\documentclass[prd,aps,floats,preprintnumbers,12pt]{article}
\usepackage{graphicx, epsfig}
\usepackage{amsmath}
\usepackage{amssymb}

\newcommand{\be}{\begin{equation}}
\newcommand{\ee}{\end{equation}}
\newcommand{\bea}{\begin{eqnarray}}
\newcommand{\eea}{\end{eqnarray}}

\begin{document}

\setlength{\unitlength}{1mm}

\begin{titlepage}

\flushright{ANL-HEP-PR-07-4}

\vspace{0.3in}

\begin{center}

{\Large \bf The Bulk RS KK-gluon at the LHC}

\vspace{0.5in}

{\bf Ben Lillie$^1$, Lisa Randall$^2$, Lian-Tao Wang$^3$}

\vspace{0.5cm}

{\it $^1$ Argonne National Lab, Argonne, IL. 60439}

\vspace{0.2cm}

{\it $\&$ University of Chicago,
  Chicago, IL 60637 }

\vspace{0.2cm}

{\it $^2$ Department of Physics, Harvard University, Cambridge MA. 02138} \\

\vspace{0.2cm}

{\it $^3$ Department of Physics, Princeton University, Princeton NJ.
08544}

\vspace{0.5cm}

\date{\today}

\vspace{0.8cm}

\end{center}

\begin{abstract}

We study the possibility of discovering and measuring the properties
of the  lightest Kaluza-Klein excitation of the gluon in a Randall-Sundrum
scenario where the Standard Model matter and gauge fields  propagate
in the bulk. The KK-gluon decays primarily into top quarks. We discuss  how to use the $t \bar{t}$ final states to
discover and probe the properties of the KK-gluon.  Identification of highly energetic tops is crucial for this analysis. We show that conventional identification methods relying
on well separated decay products will  not work  for heavy resonances but suggest alternative methods for top identification for  energetic tops.
We find,
conservatively, that resonances with masses less than $5$ TeV can be
discovered if the algorithm to identify high $p_T$ tops can reject
the QCD background by a factor of 10. We also find that for similar
or lighter masses the spin  can be determined and
for lighter masses  the chirality of the coupling to $t\bar
t$ can be measured. Since the energetic top pair final state is a
generic signature for a large class of new physics as the top
quark presumably couples most strongly to the electroweak symmetry
breaking sector, the methods we have outlined to study the
properties of the KK-gluon should also be important in other
scenarios.
\end{abstract}
\end{titlepage}

\section{Introduction}
Proposed higher-dimensional solutions to the hierarchy
problem suggest novel experimental possibilities for the LHC.
An interesting class of models falls in the  so-called Randall-Sundrum scenario in with a single warped extra dimension
\cite{RS}. In the original model the Standard
Model (SM) fields were confined to a single brane. However
models in which the SM fermions and gauge bosons
propagate throughout all five dimensions
\cite{RSmatterinbulk,RSflavor,RSmodel} can have  attractive features
 such as suppressing flavor-changing neutral currents, explaining
the hierarchy of fermion masses \cite{RSflavor}, and allowing gauge
coupling unification \cite{RSunification}.

The first signal of such a framework should be the resonant
production of the Kaluza-Klein (KK) excitations of the gauge bosons.
Since the KK-gluons are the most strongly coupled of the new KK particles, they  have the
largest production rate and will likely be the first signal of this
model. We study the potential reach for discovery of this state in this paper.

As we will see below, all gauge KK states decay primarily to
$t\bar t$, so the KK modes of the electroweak bosons do not gain in
signal significance by having cleaner decay channels. Although
discovering the KK-gluon will not in itself definitively establish
the underlying RS geometry,  a generic composite model  of
electroweak symmetry breaking does not require the existence of a
KK-gluon. Therefore observing such a state could be interpreted as evidence  for an extra-dimensional scenario to be supplemented by further studies.
Although it is certainly interesting, for example,  to
study decay modes such as $t^{(1)} \rightarrow g^{(1)} + t$, we
focus here on the more generic decay mode $g^{(1)} \rightarrow t
\bar{t}$.

The masses of KK fermions are generically comparable to
the KK-gluons, so whether the decay of the gluon KK modes to fermion
KK states is open or not is  model-dependent. In order to
account for the top quark mass, the top quark is likely to be
localized towards the TeV brane or mostly composite as in the model
of, for example,  Ref.~\cite{RSmodel}. Since the KK-gluon is also localized towards the TeV brane, it has a
large wavefunction overlap with other composite states so we expect the KK-gluon decays
dominantly into $t \bar{t}$ final states. Since  the mass
of the KK-gluons should be $\mathcal{O}(\mbox{TeV})$, their decays give
energetic top quark final states, which pose novel challenges to
experimental searches.

In this paper we  first study the discovery reach for a KK-gluon. We
then study the determination of its spin assuming decay into highly
energetic top quark final states. Efficient identification of the top quark
final states will require the implementation of new experimental
methods. The conventional method for top jet identification based on reconstruction from
separate objects in the top decay (see, for example,
\cite{top_convention,atlasTDR,cmsTDR} ) won't work for much of the
paramater range since the tops will be highly boosted. New
approaches, such as identifying massive jets, merit careful
investigation. In this paper we point out the importance of such a
signal and consider the necessary degree of top-jet resolution.
Detailed knowledge of the structure of jets as well as sophisticated
detector simulation will be needed to determine the efficacy of this
approach, which is beyond the scope of this paper. Therefore, rather
than focusing on a specific strategy, we choose to parameterize
these unknowns  with a range of signal dilution factors to determine
the discovery potential in each case. We assume this will be followed
by further experimental study.

We will also consider the possible
spin-determination efficiency as well as the question of the
helicity of the top final state. This is particularly interesting
because in the simplest implementation \cite{RSmodel} of a composite
top, the $t_R$ is localized towards the TeV brane while $(t_L,b_L)$
is almost flat, and therefore the top quarks from KK-gluon decays
will be highly polarized.

\section{Couplings between fermions and the KK gluon}

The RS construction is a slice of 5 dimensional anti-deSitter space,
$AdS_5$, with metric
\begin{gather}
ds^2 = e^{-2\sigma}\eta_{\mu\nu}dx^\mu dx^\nu - dy^2
\end{gather}
where $y$ is the coordinate of the 5th direction, and $\sigma = k|y|$
with $k$ determining the curvature of the $AdS_5$. The extra
dimension has the geometry of the orbifold $S_1/Z_2$. At either
fixed point of the orbifold is a brane. One, at $y=0$ is called the
``Planck'', or ``UV'' brane; the other, at $y = \pm\pi r_c$, is the
``TeV'' or ``IR'' brane. If $kr_c \simeq 11$ then the warp factor at
the IR brane, $\epsilon = e^{-\pi k r_c} = 10^{-15}$ is the ratio of
the Planck to TeV scales\footnote{For definiteness, we take $kr_c$ =
11.27 to match to previous studies.}.

The couplings between the SM fermions and the KK-gauge bosons, in particular
the KK-gluon, determine most of the interesting phenomenological
features of the model. We provide a brief summary of them here. See
Appendix~\ref{formalisms} for a detailed discussion of usefulh
formalisms for describing gauge and fermion fields in $AdS_5$.

\begin{figure}
\begin{tabular}{cc}
\includegraphics[angle=270,scale=0.3]{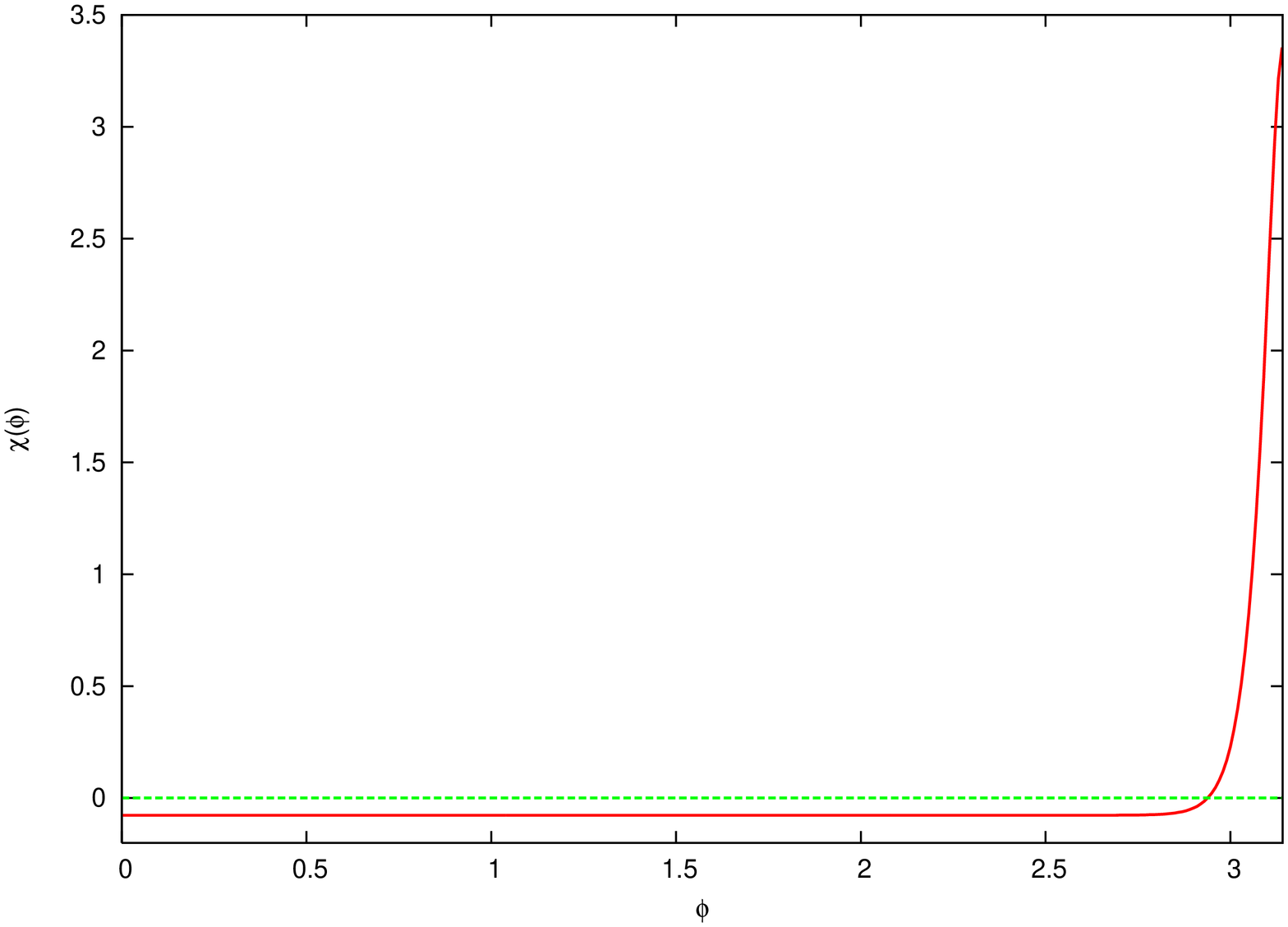} &
\includegraphics[angle=270,scale=0.3]{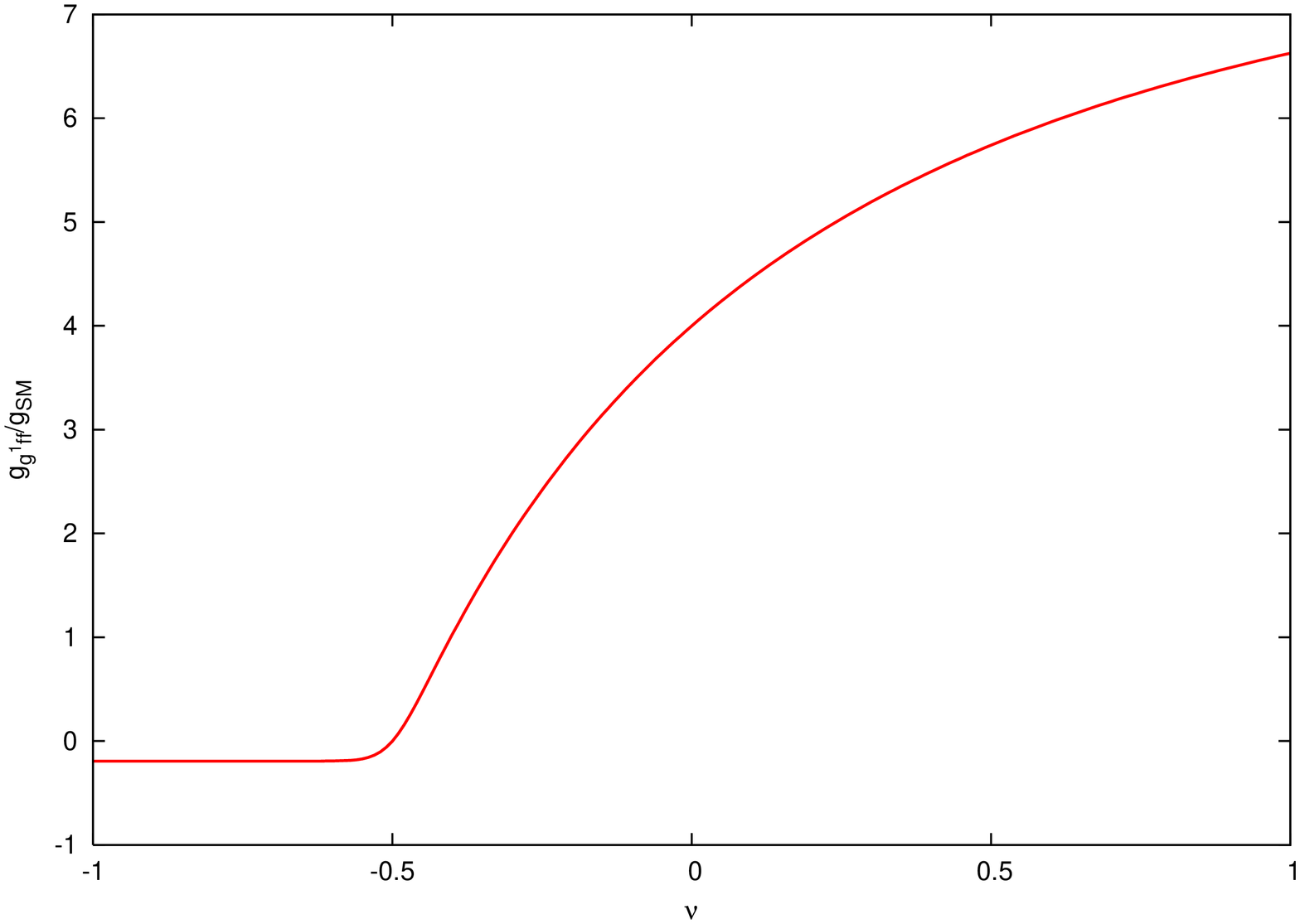} \\
\end{tabular}
\caption{\label{fig:gffcoup}Left: Wavefunction of the first excited
gauge KK mode. Note that it is extremely flat near the UV brane,
with $AdS_5$ bulk coordinate defined by $y=\pi \phi$. Right:
Coupling of the first gauge KK mode to a zero mode fermion as a
function of the fermion localization parameter, $\nu$. Notice that
this coupling asymptotes quickly to a constant value for $\nu <
-0.5$}
\end{figure}

We obtain the 4D effective couplings $g_{f\bar f A^{(n)}}$ by
integrating over the 5th dimension. These couplings are
sensitive to the profiles of the fermion and KK-gauge boson
states in the bulk. Modulo certain brane kinetic terms, which we are
not considering here, the profile of the KK-gluon is largely fixed,
as can be seen in the left panel of Fig.~\ref{fig:gffcoup}. On the
other hand, the wavefunctions of zero mode fermions are model
dependent, and can be used to address the flavor hierarchy.

For a 5D fermion sector, we can write down a mass term
$m_{\Psi}\bar\Psi\Psi$ for each 5d fermion field. Since $\Psi$ is
Dirac, we can write it as $\Psi = (\zeta\ \xi)^\top$ where $\zeta$
and $\xi$ are Weyl spinors. We can write the mass as $m_{\Psi} =
\nu_{\Psi}k \epsilon(y)$, where $\epsilon(y)$, which is $1$ for
$y>0$ and $-1$ for $y<0$, is responsible for making the mass term
even under the orbifolding symmetry. The orbifold projection selects
out one zero mode of one chirality. The wavefunction of the
remaining zero mode is \cite{RSflavor}
\begin{gather}
\xi_{\Psi}^{(0)} = \frac{e^{\nu_{\Psi} k y}}{N_f^{(0)}},
\end{gather}
where $N_f^{(0)}$ is a normalization factor. This wavefunction is
exponentially peaked  toward the UV brane for $\nu < -1/2$ and
toward the IR for $\nu > -1/2$.

Assuming $\mathcal{O}(1)$ 5D Yukawa couplings, the effective 4D
Yukawa couplings are determined by the wavefunction overlap between
the SM fermions and the TeV brane. Since the top quark Yukawa is very
close to unity, the 3rd generation cannot be strongly UV localized.
However, localizing the $b_L$ (which, of course, comes in a doublet
with the $t_L$) too close to the IR brane results in unacceptable
deviations in the rate of $Z\to b\bar b$ \cite{RSmodel}. This leads
to the following set of parameters
\begin{align}
\nu_{t_R} &\approx -0.3,\notag\\
\nu_{Q_{3L}} & \approx -0.4,\label{eq:fermions}\\
\nu_{\rm other} & < -0.5.\notag
\end{align}

That is: the $t_R$ is IR localized, the third generation quark
doublet, $Q_{3L}$ is close to flat, and the others are all $UV$
localized. This is important because the coupling of a zero mode
fermion to a KK-gauge  boson is dependent on the $\nu$ parameter.
This coupling is given by\footnote{See the appendix for definitions
of the symbols.} \cite{RSmatterinbulk}
\begin{gather}
g_{f\bar f A^{(n)}} = g_4 \sqrt{2\pi k r_c}
\left[\frac{1+2\nu}{1-\epsilon^{2\nu+1}}\right]
\int_{\epsilon}^{1}dz z^{2\nu+1} \frac{J_1(x_n^A z) + \alpha_n^A
Y_1(x_n^A)}{|J_1(x_n^A)+\alpha_n^A Y_1(x_n)^A|}.
\end{gather}
The right panel of Fig. \ref{fig:gffcoup} shows the ratio of
couplings, $g_{ffA^{(1)}}/g_4$, for the first gauge KK mode. Note
that for $\nu < -0.5$ the coupling is flat in $\nu$ and
$g_{ffg^{(1)}} \simeq 0.2 g_4$, where the effective 4D gauge
coupling of the $SU(3)_C$ gauge coupling in the Standard Model, $g_4
= g_s$. Notice that the production cross-section for the process $q
\bar{q} \rightarrow g^{(1)}$ is not suppressed by $M_P$
\cite{Pomarol:2000hp}, as it is for the KK gravitons. This is
generically true for any KK-resonances which are not singlets of the
SM gauge symmetry. This fact can also be understood in the CFT dual as
the effect of mixing between fundamental and composite states
\cite{adscft}. For the $Q_{3L}$ doublet we have $g_{ffg^{(1)}} \simeq
g_s$, and for the $t_R$, $g_{ffg^{(1)}} \simeq 4g_s$.

The fact that the coupling constant  $g_{f \bar{f}A^{(n)}}$ approaches
a constant results from near flatness of $A^{(n)}$ wave-function (as
shown in Fig.~\ref{fig:gffcoup}) and
the highly localized nature of the fermion wavefunction for $\nu <
-0.5$ ($\propto \exp (-(1+2 \nu) y)$ away from the Planck
brane). Using the asymptotic form of the $A^{(n)}$ wave-function, we
estimate  from Eq.  4 \cite{RSmatterinbulk}:
\begin{equation}
\frac{g_{f \bar{f} A^{(n)}}}{g_{\rm SM}} \simeq - \frac{1}{\sqrt{\pi k
    r_c}} \frac{1}{\sqrt{n - 0.2} }  +  {\mathcal{O}}  [(\pi k
  r_c)^{-3/2}] .
\end{equation}
The fact that this coupling is not exponentially suppressed
\cite{RSmatterinbulk} can be understood in the CFT picture
as the result of kinetic mixing between the SM gauge field and the
composite vector states (KK-gauge bosons in the AdS) \cite{adscft}.

Note that we are looking only at KK-gluon production from
quarks. At tree-level the coupling of two zero-mode gluons to a KK
mode is zero by orthogonality. Though there may be a loop-induced
coupling, the main effect of this would be to increase the rate,
since the $gg$ channel will open. Neglecting it here is therefore a
conservative assumption.

Constraints on the masses of the KK excitations of the
electroweak gauge bosons from precision electroweak data give
$M_{KK} \gtrsim 2-3$ TeV \cite{RSmodel}. However, such a constraint
is model-dependent. Since we are interested in generic properties of
the KK-gluon, we  present our result in a broader range of the
KK-gluon mass.

\section{Discovery}

\subsection{Cross Section}

\begin{figure}
\begin{center}
\includegraphics[angle=270,scale=0.5]{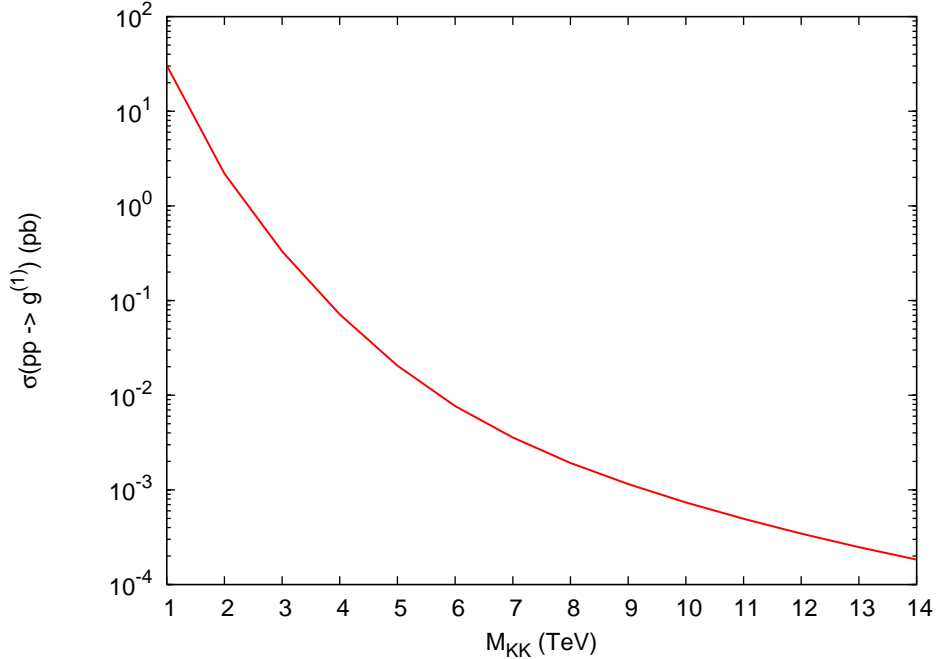}
\caption{\label{fig:xsec} Total cross-section for production of
  the first KK gluon, as a function of KK mass. }
\end{center}
\end{figure}

The KK excitations of the gluons  will appear as
resonances in the process $pp \to q\bar q$, primarily decaying in the  $t\bar t$ channel.   The branching ratio for $g^{(1)} \to t\bar t$ is $92.5\%$
(and another $5.5\%$ is to $b\bar b$, with the rest to light quark
jets). To study the signal we have simulated the process $q\bar q
\to g^{(1)} \to q\bar q$ using MADGRAPH and MADEVENT
\cite{Maltoni:2002qb}. A plot of the inclusive cross section as a
function of the resonance mass is shown in Fig. \ref{fig:xsec}. The
width of this resonance with the fermion configuration in Eq.
\ref{eq:fermions} is
\begin{gather}
\Gamma/M \approx 0.17.
\end{gather}

\begin{figure}
\begin{center}
\includegraphics[angle=270,scale=0.5]{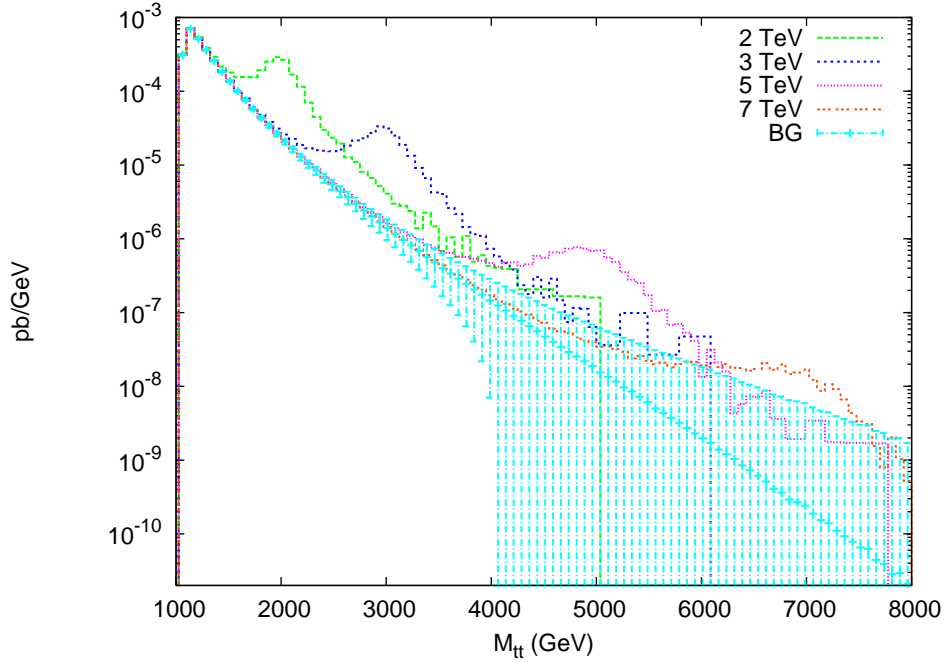}
\end{center}
\caption{\label{fig:mtt}Invariant mass distribution of $t\bar t$
pairs coming from the KK gluon resonance, and SM $t\bar t$
production. The errors shown on the background curve are the
statistical errors assuming 100 $fb^{-1}$ of luminosity.}
\end{figure}

Figure \ref{fig:mtt} shows the $t\bar t$ invariant mass distribution
from KK resonances, demonstrating  that with efficient top quark
identification it should be visible above the SM $t\bar t$
background up to relatively high masses. This will require
reconstructing the $t \bar{t}$ pair to identify the relatively
narrow resonance in the $m_{t \bar{t}}$ distribution. Clearly,
identifying the top pairs will be  crucial to the discovery and
 study of the KK-gluon and experiments will have to be as efficient as possible in identifying tops.

\begin{figure}
\begin{center}
\includegraphics[angle=270,scale=0.5]{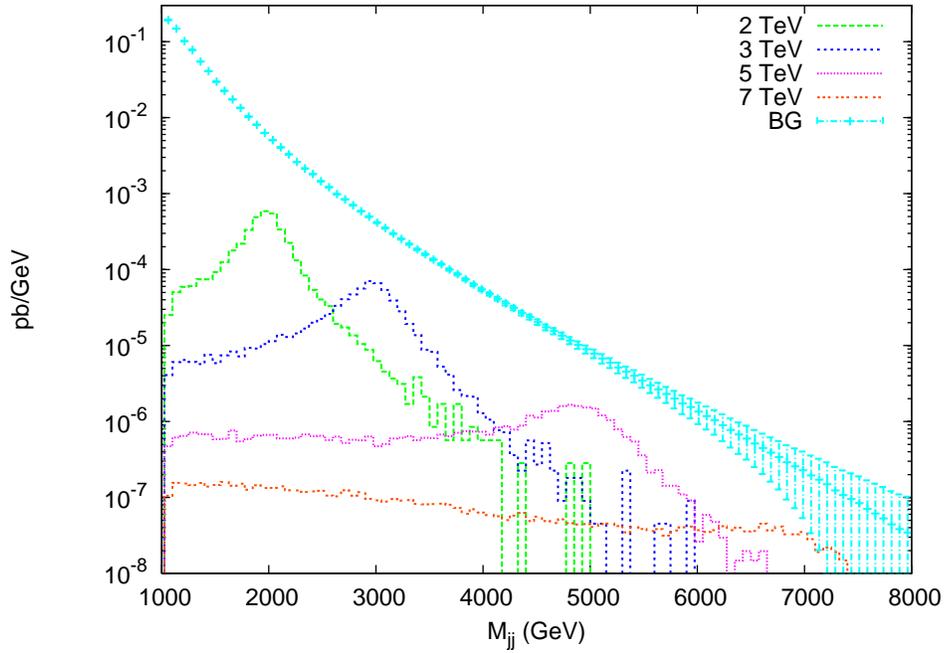}
\end{center}
\caption{\label{fig:mjj} Invariant mass distribution of the  decay
products for several masses of the KK gluon. This assumes all $t\bar
t$ events are fully collimated. ``BG'' is QCD dijet production. All
jets are required to have pseudo-rapidities $|\eta| < 0.5$, and at
least one to have $p_T > 500$ GeV. The errors shown on the
background curve are the statistical errors assuming 100 $fb^{-1}$
of luminosity.}
\end{figure}

To emphasize the importance of top ID,  consider
the worst case scenario in which a top jet is not distinguished from
a QCD jet. We compare the signal with QCD dijet production.  We show the rates for dijets, with both
pseudo-rapidities $< 0.5$ and the leading jet $p_T > 500$ GeV in
Fig. \ref{fig:mjj}.
  We see that even selecting the events to be very
central and containing high $p_T$ jets, signal identification is
difficult. The raw dijet rate is overwhelming
even with these cuts. Although more refined cuts could reduce the
background, they are probably not enough without some top-quark ID.

So before further further discussing
   the KK gluon resonance, we
consider  collider features of energetic top quarks decaying
from the resonance in the next subsection.

We note that in principle $A^{(1)} \rightarrow W^+_L W^-_L$ is also
an important and generic decay modes for KK states, such as the KK
states of the $Z$ or gravitons, due to the large overlap between
their wavefunctions and the longitudinal mode of $W$. However, such
a decay mode does not exist for KK-gluon. In fact, we could treat
such an absence as one other confirmation that the resonance is a
KK-gluon.


\subsection{Characteristics of energetic tops}

 In this section we discuss possible top quark
collider signatures.

Top reconstruction in $t \bar{t}$ final states has been very well
studied for relatively low invariant mass, $m_{t \bar{t}} \leq 1$
TeV. A crucial ingredient of the algorithm is detecting several
separated objects from the decaying tops
\cite{top_convention,atlasTDR,cmsTDR,top-tao}. For example, in the
promising semi-leptonic decay, typically an  isolated
lepton and a $b$-jet on one side and three jets on the other side are required.
However, we expect top quarks from $g^{(1)}$ decays to be very
energetic, with $p_T \sim 0.5 M_{g^{(1)}}$. Therefore, for such
energetic top quarks (especially at the high end of the mass range)
we expect the decay products will be highly collimated and the
conventional reconstruction algorithm will fail.

\begin{figure}
\begin{center}
\includegraphics[angle=270,scale=0.3]{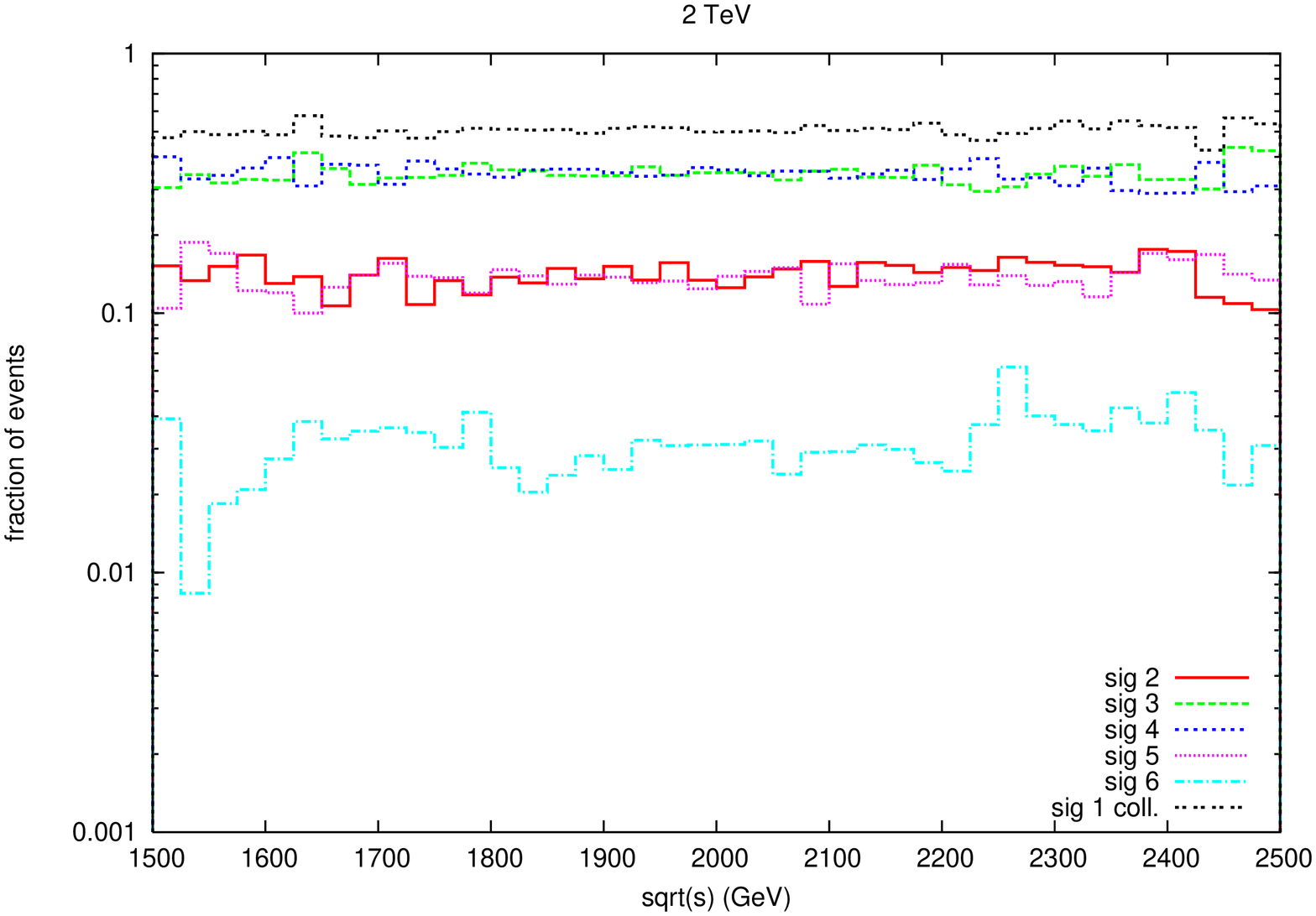}
\includegraphics[angle=270,scale=0.3]{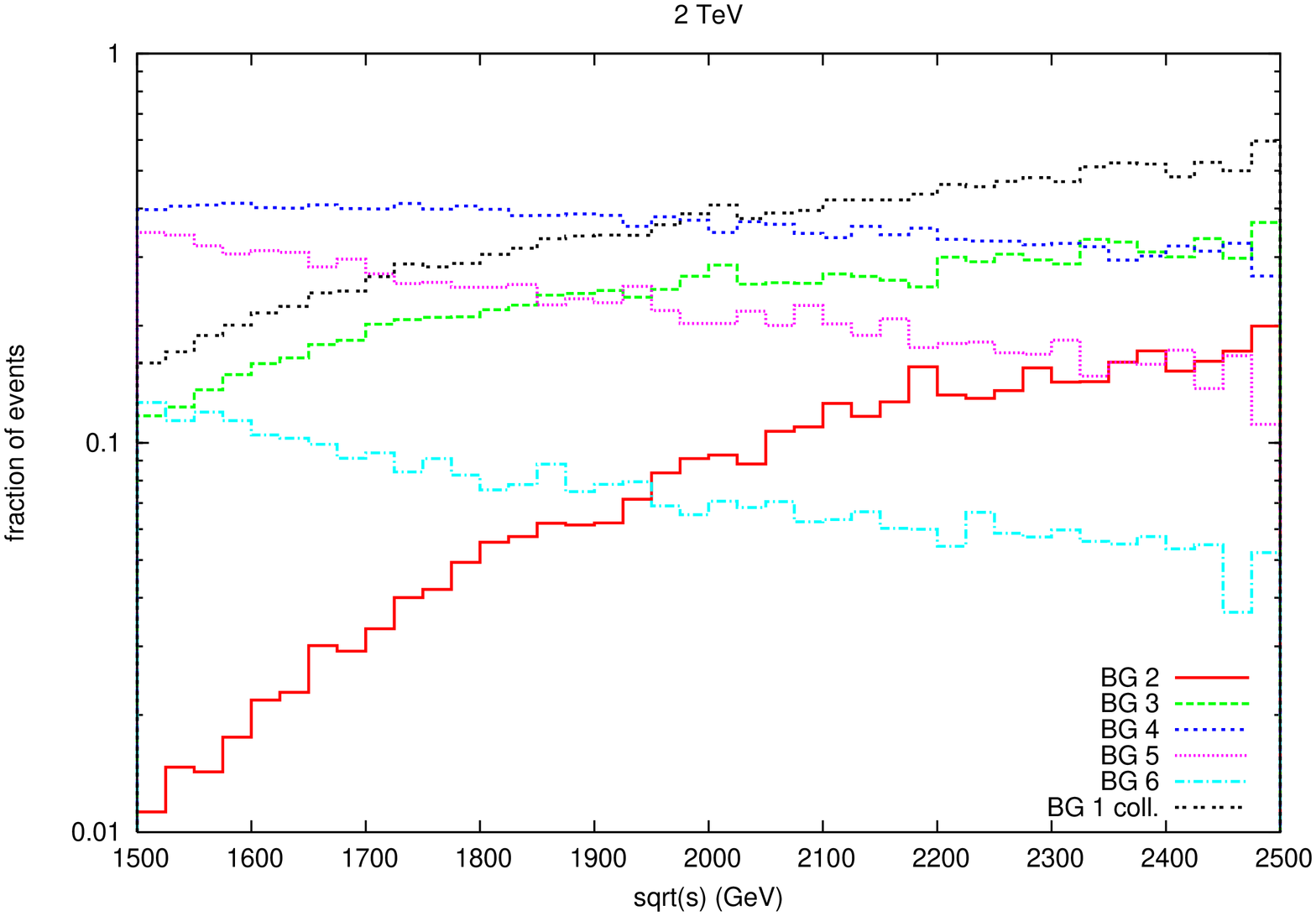}
\includegraphics[angle=270,scale=0.3]{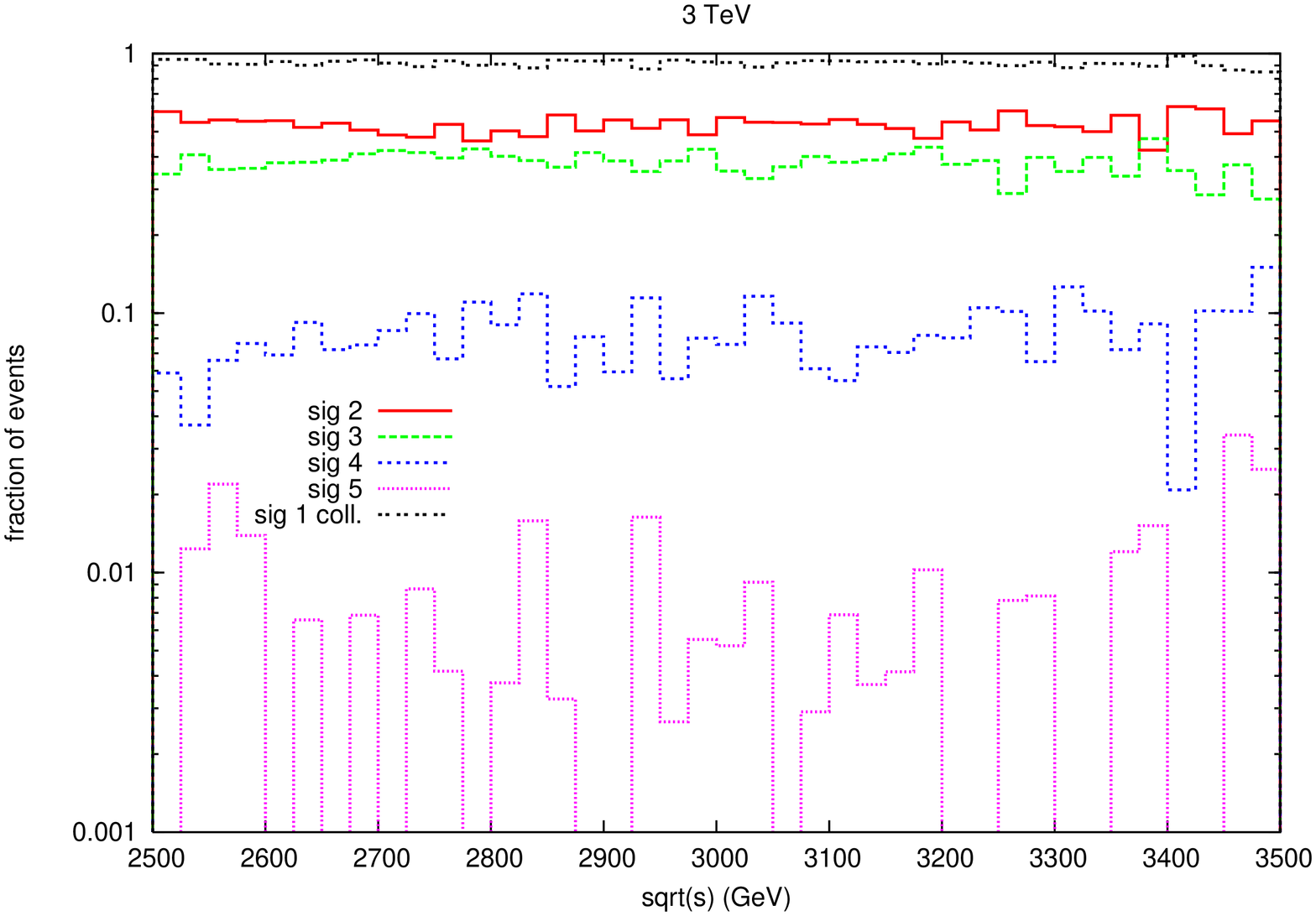}
\includegraphics[angle=270,scale=0.3]{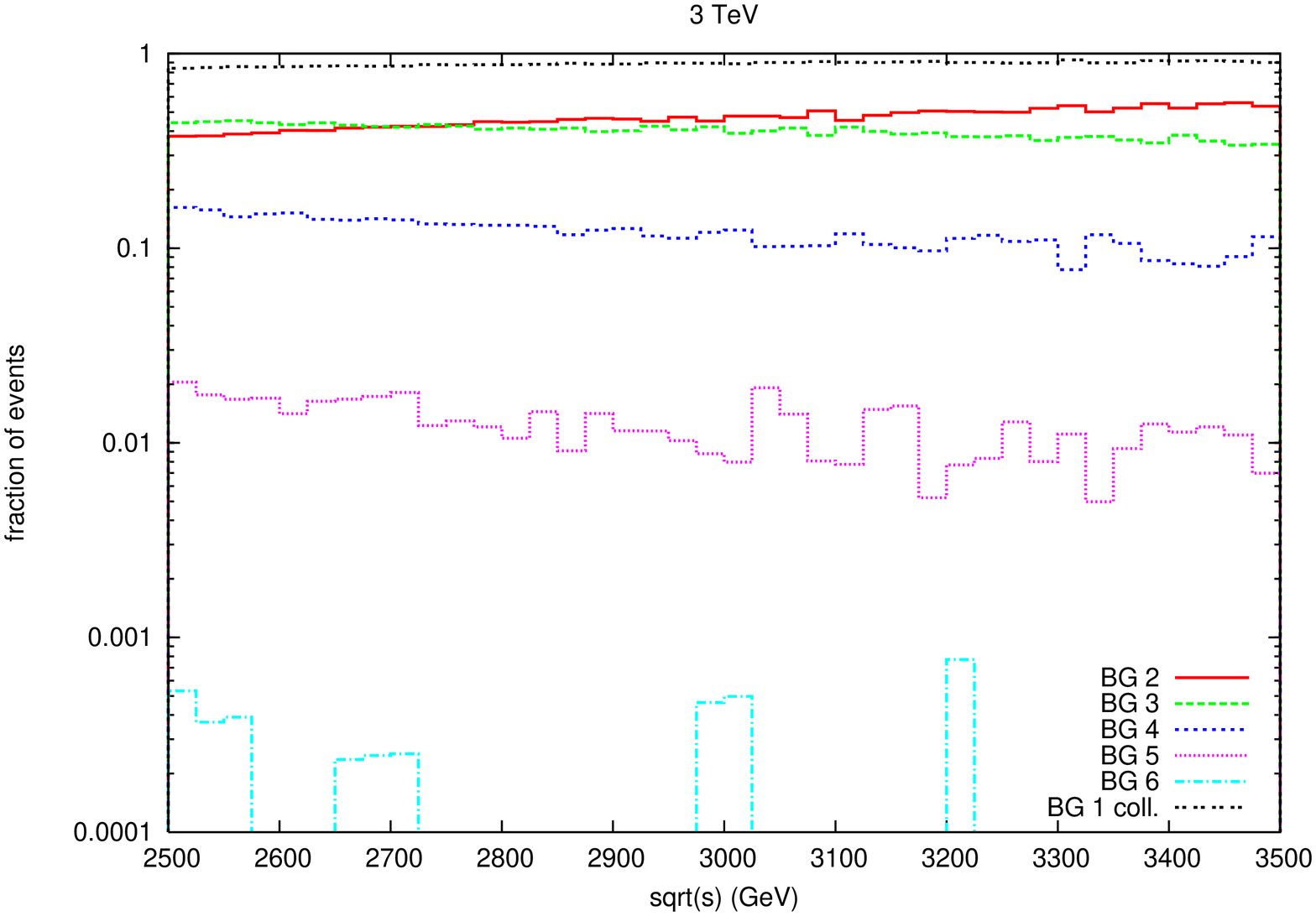}
\includegraphics[angle=270,scale=0.3]{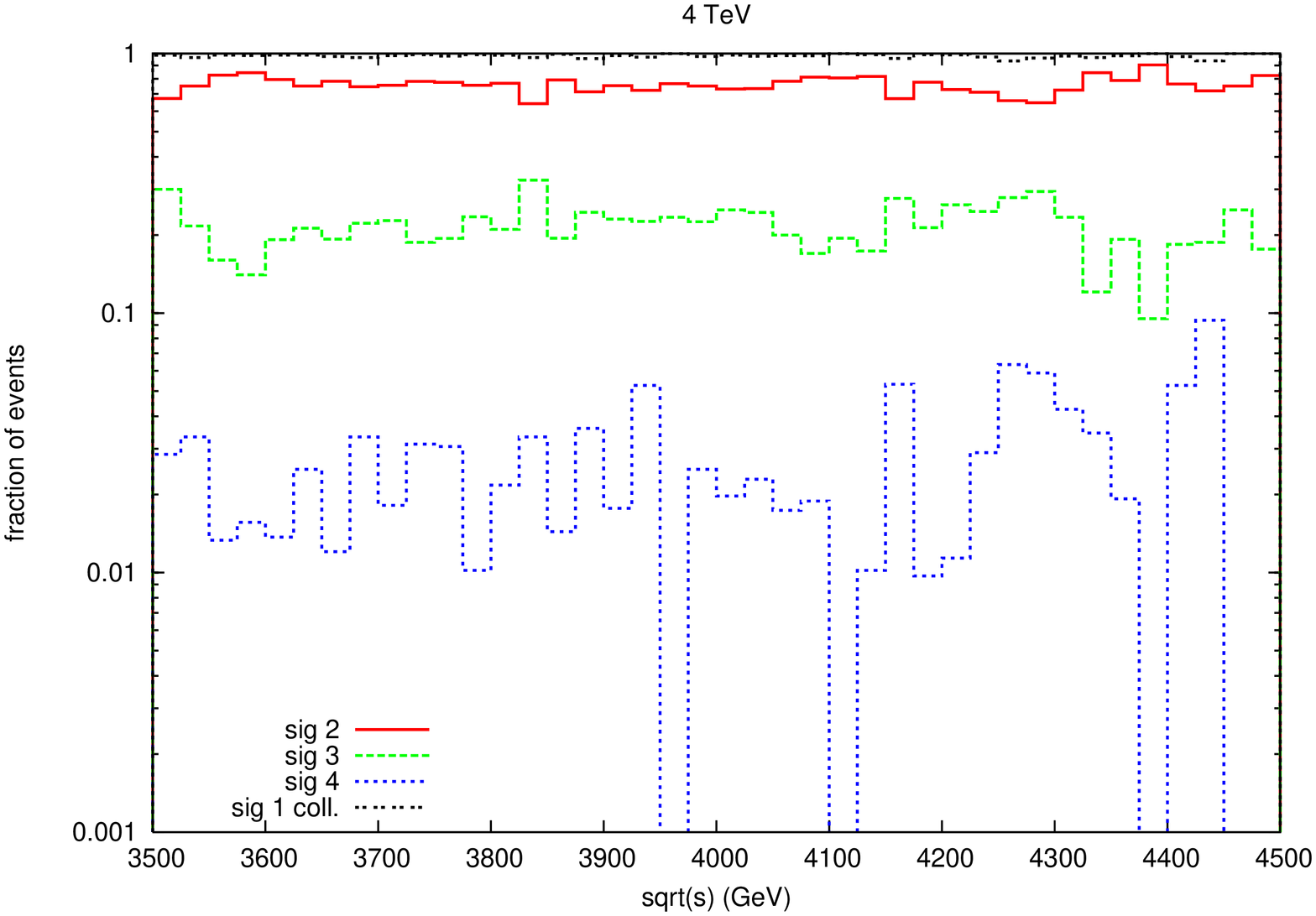}
\includegraphics[angle=270,scale=0.3]{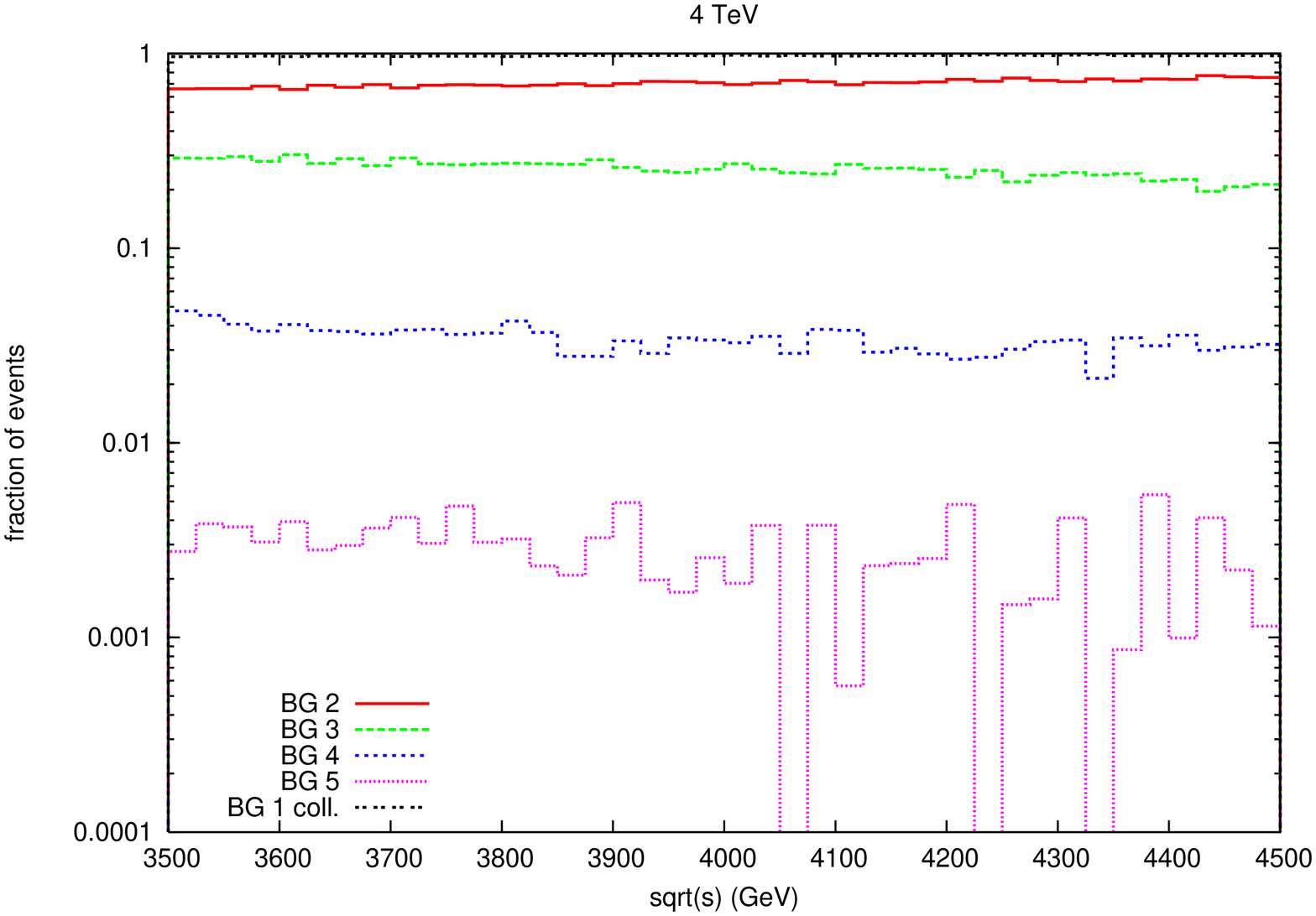}
\end{center}
\caption{\label{fig:nobj}Left: Fraction of events with certain
numbers of distinct objects for events from decay of a KK gluon,
with mass (top to bottom) 2, 3, and 4 TeV as a function of invariant
mass of the $t\bar t$ pair, after imposing a cut on the top $p_T$:
500 GeV, 1 TeV, and 1.5 TeV. Right: SM $t\bar t$ production using
the same cuts as the corresponding plot on the right. The line
labeled ``1 coll." is the fraction of events where at least one of
the tops has all three decay products within the same cone. A cone
size of 0.4 has been used.}
\end{figure}

\begin{figure}
\begin{center}
\includegraphics[angle=270,scale=0.3]{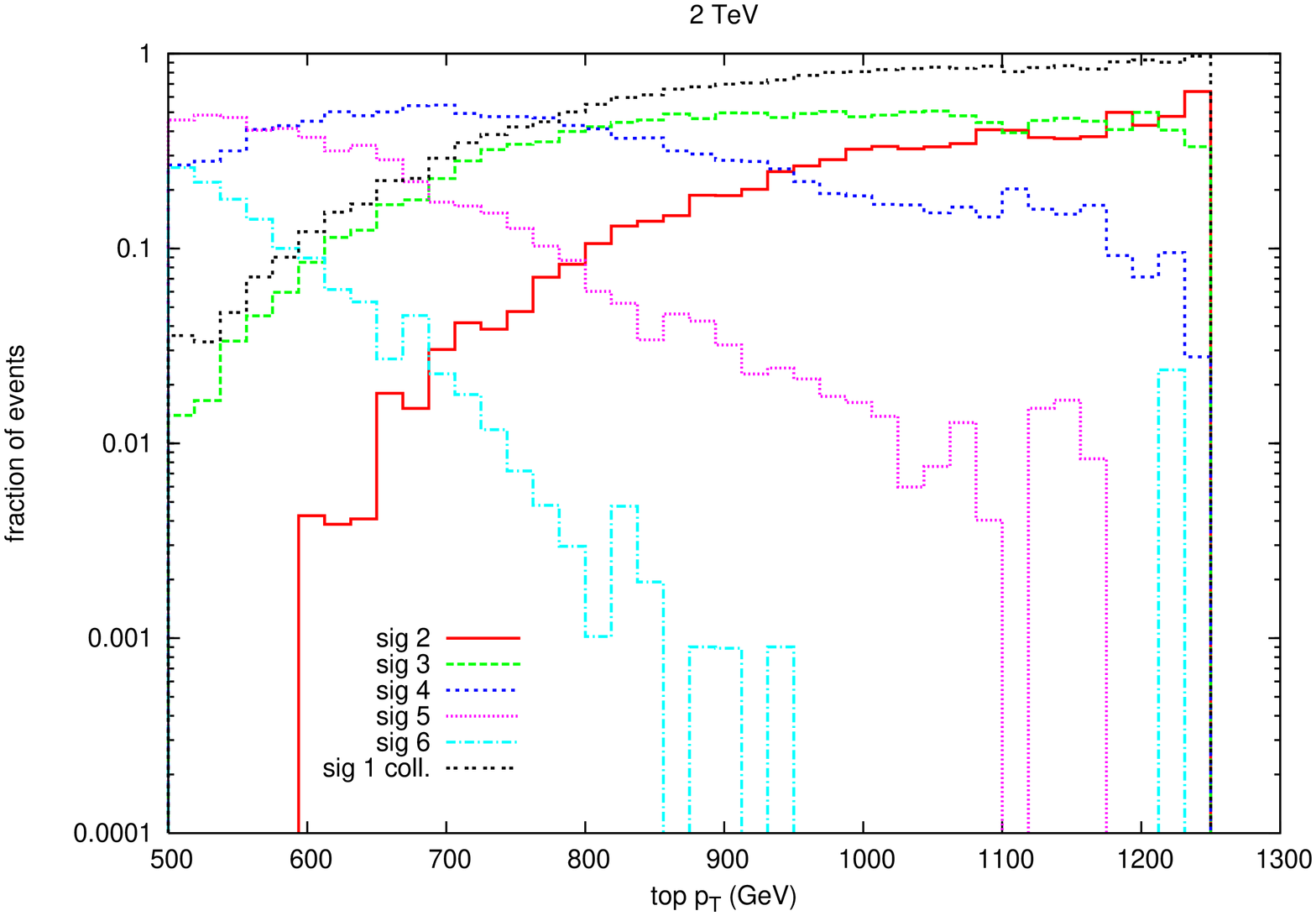}
\includegraphics[angle=270,scale=0.3]{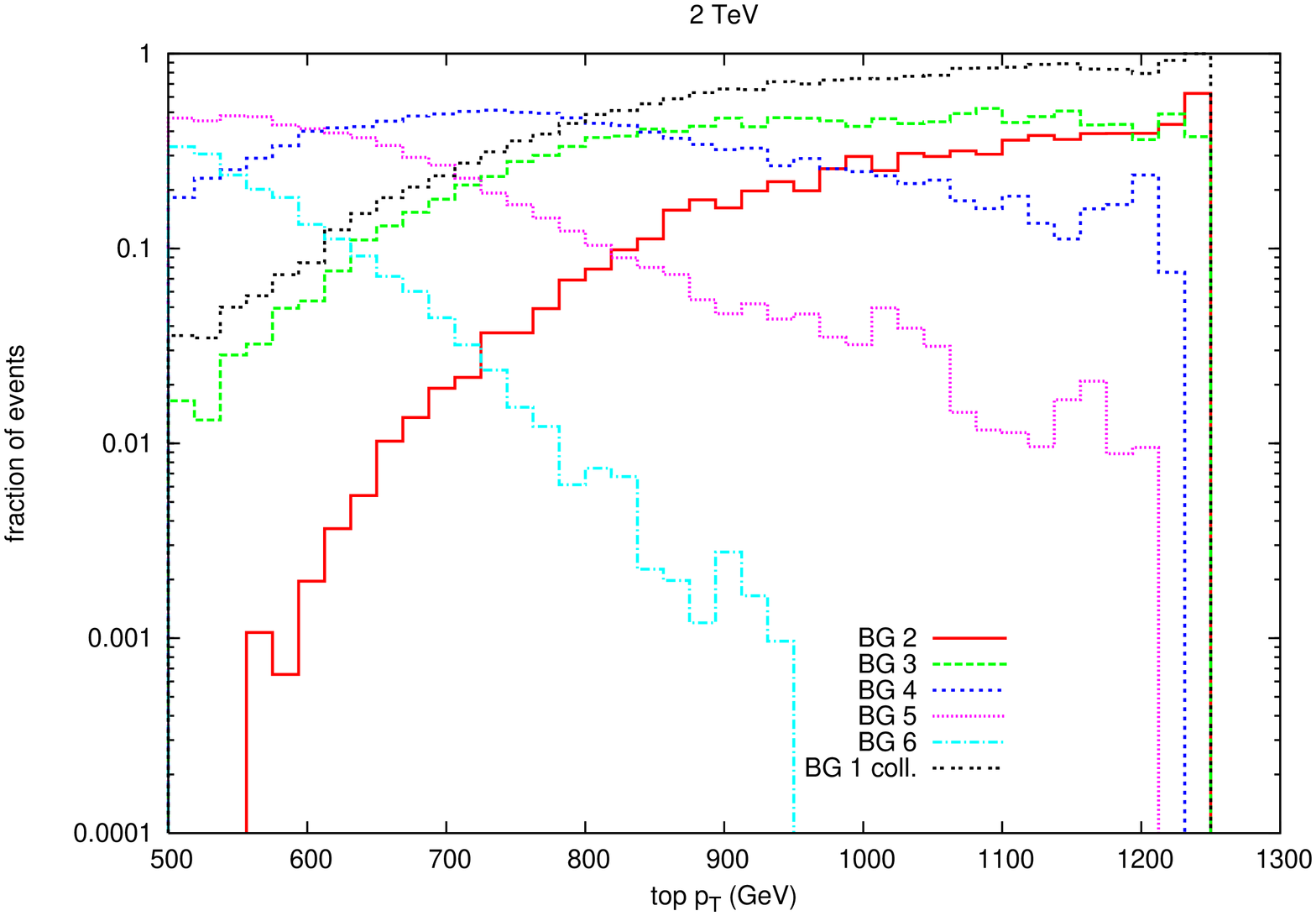}
\includegraphics[angle=270,scale=0.3]{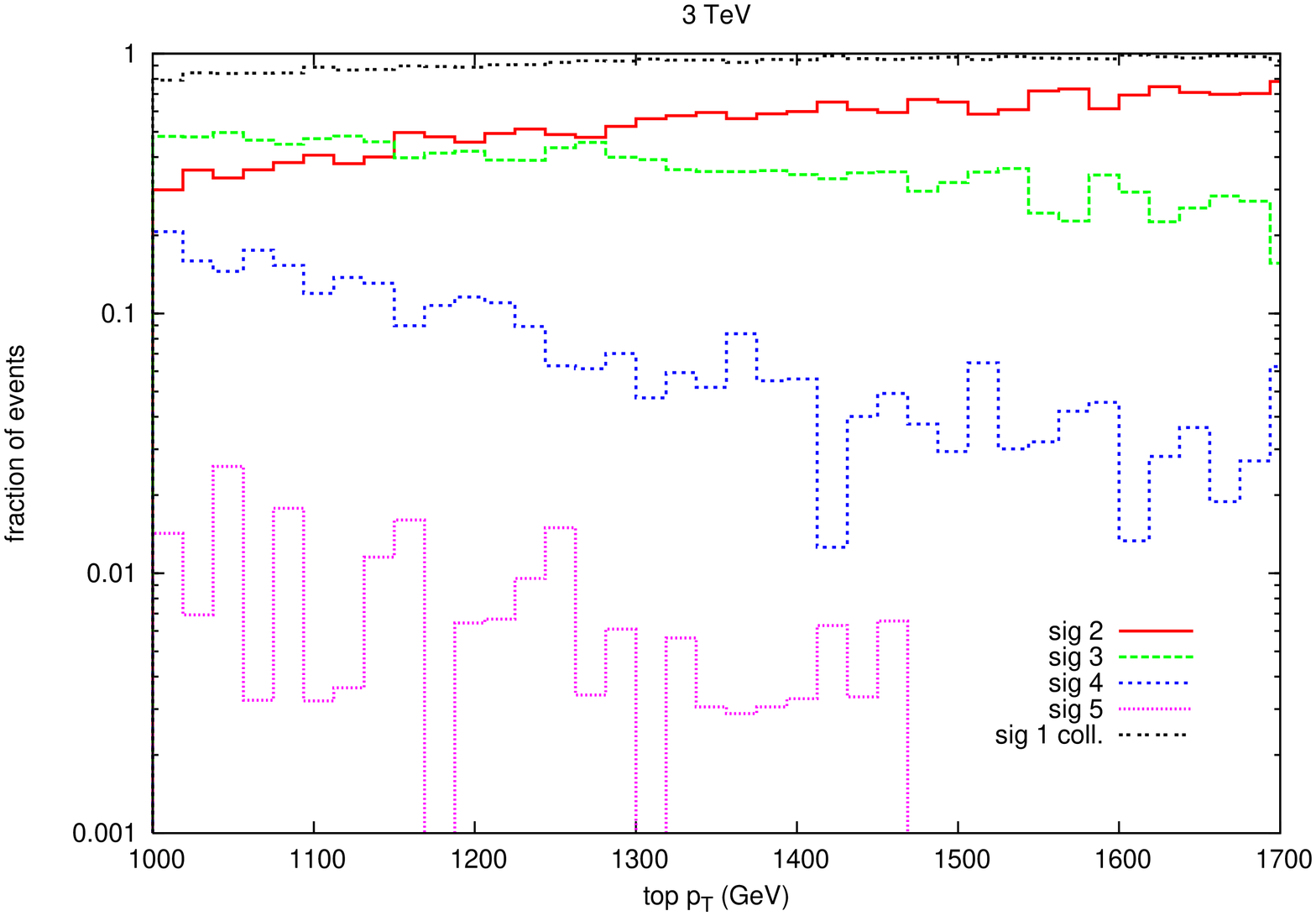}
\includegraphics[angle=270,scale=0.3]{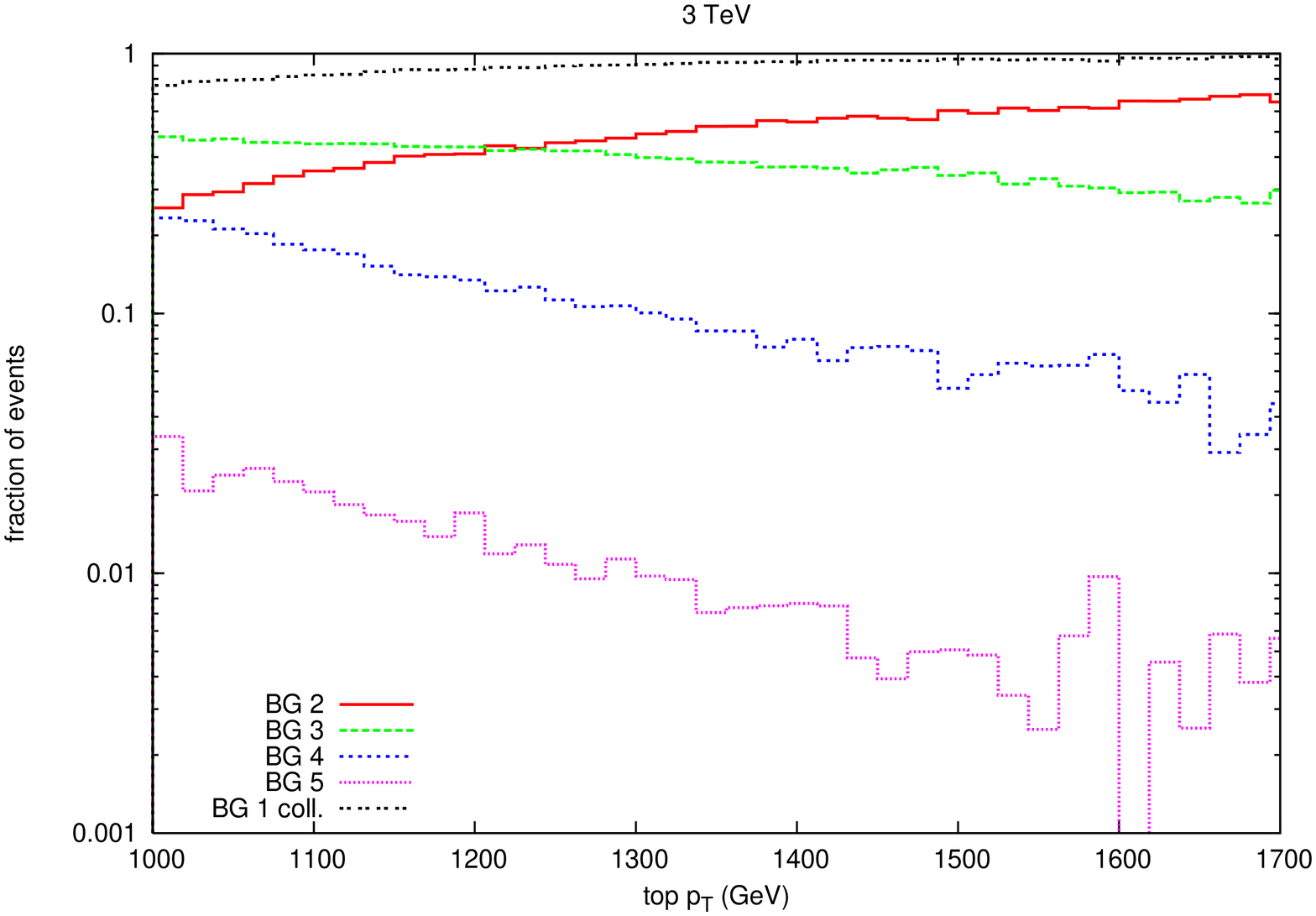}
\includegraphics[angle=270,scale=0.3]{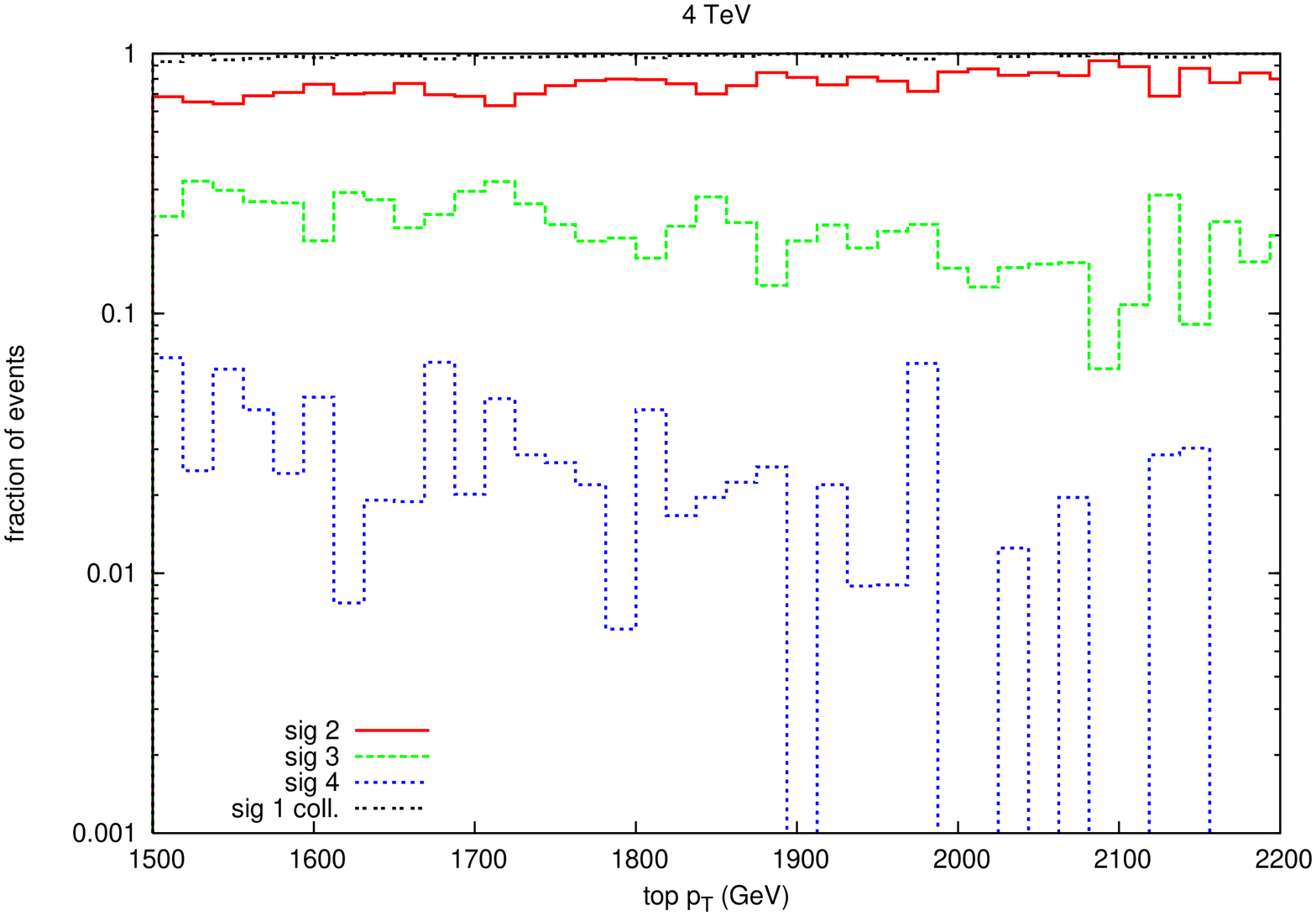}
\includegraphics[angle=270,scale=0.3]{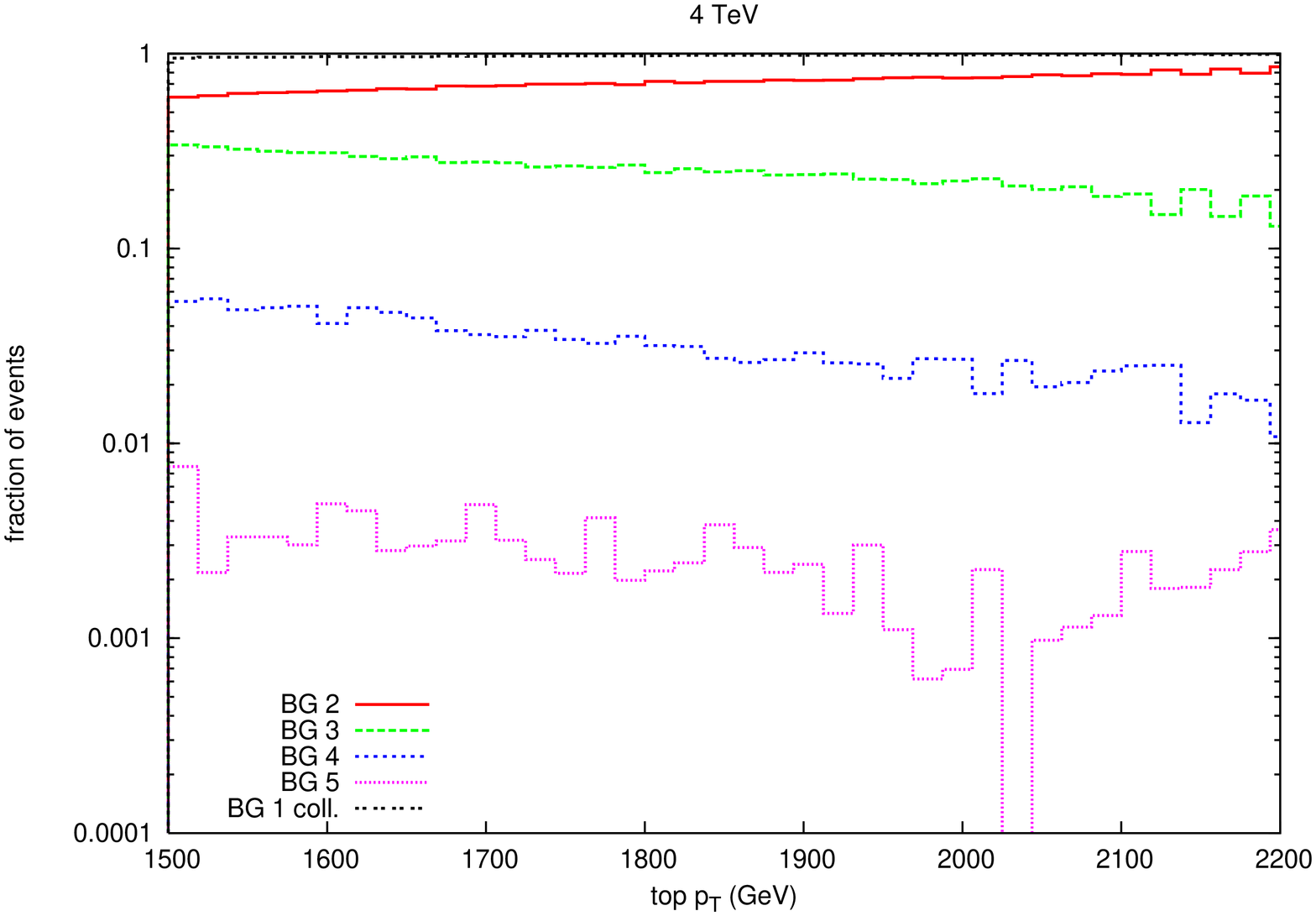}
\end{center}
\caption{\label{fig:nobj-pt}Left: Fraction of events for certain
numbers of distinct objects for events from decay of a KK gluon,
with mass (top to bottom) 2, 3, and 4 TeV as a function of $p_T$ for
events in the window $m_{KK} - 500{\ \rm GeV} < m_{tt} < m_{KK} +
500$ GeV. Right: SM $t\bar t$ production using the same cuts as the
corresponding plot on the right. The line labeled ``1 coll." is the
fraction of events where at least one of the tops has all three
decay products within the same cone. A cone size of 0.4 has been
used.}
\end{figure}

\begin{figure}
\begin{center}
\includegraphics[angle=270,scale=0.3]{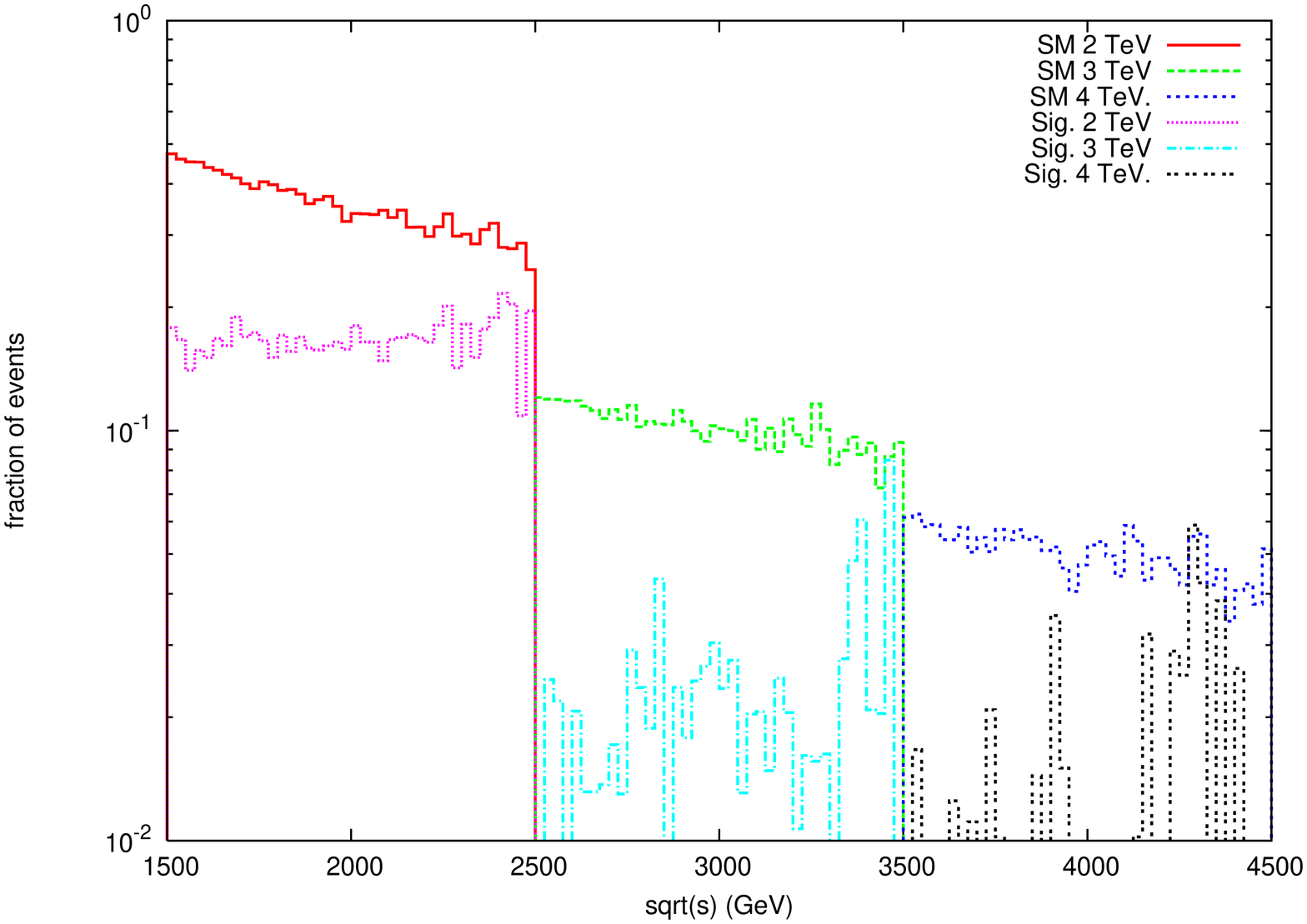}
\includegraphics[angle=270,scale=0.3]{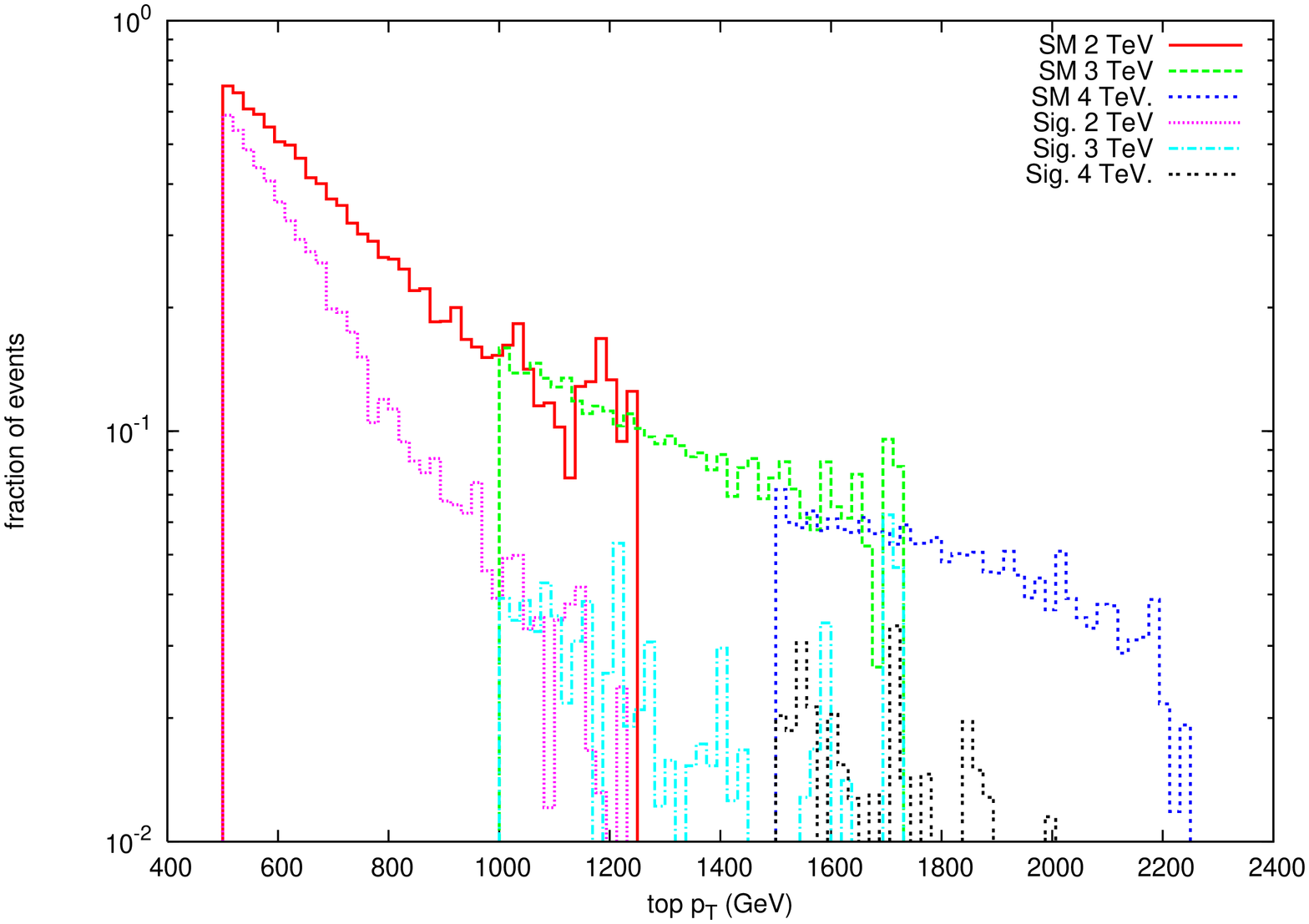}
\end{center}
\caption{\label{fig:leps}Left: Fraction of events where at least 1
top decays leptonically with an isolated lepton. The cuts imposed
are: top $p_T
>$ (500, 1,000, 1,500) GeV for KK masses (2,3,4) TeV, the invariant
mass in a window 1 TeV wide around the resonance. The lines labeled
"SM" are for SM $t\bar t$ production with the same cuts.}
\end{figure}

\begin{figure}
\begin{center}
\includegraphics[angle=270,scale=0.5]{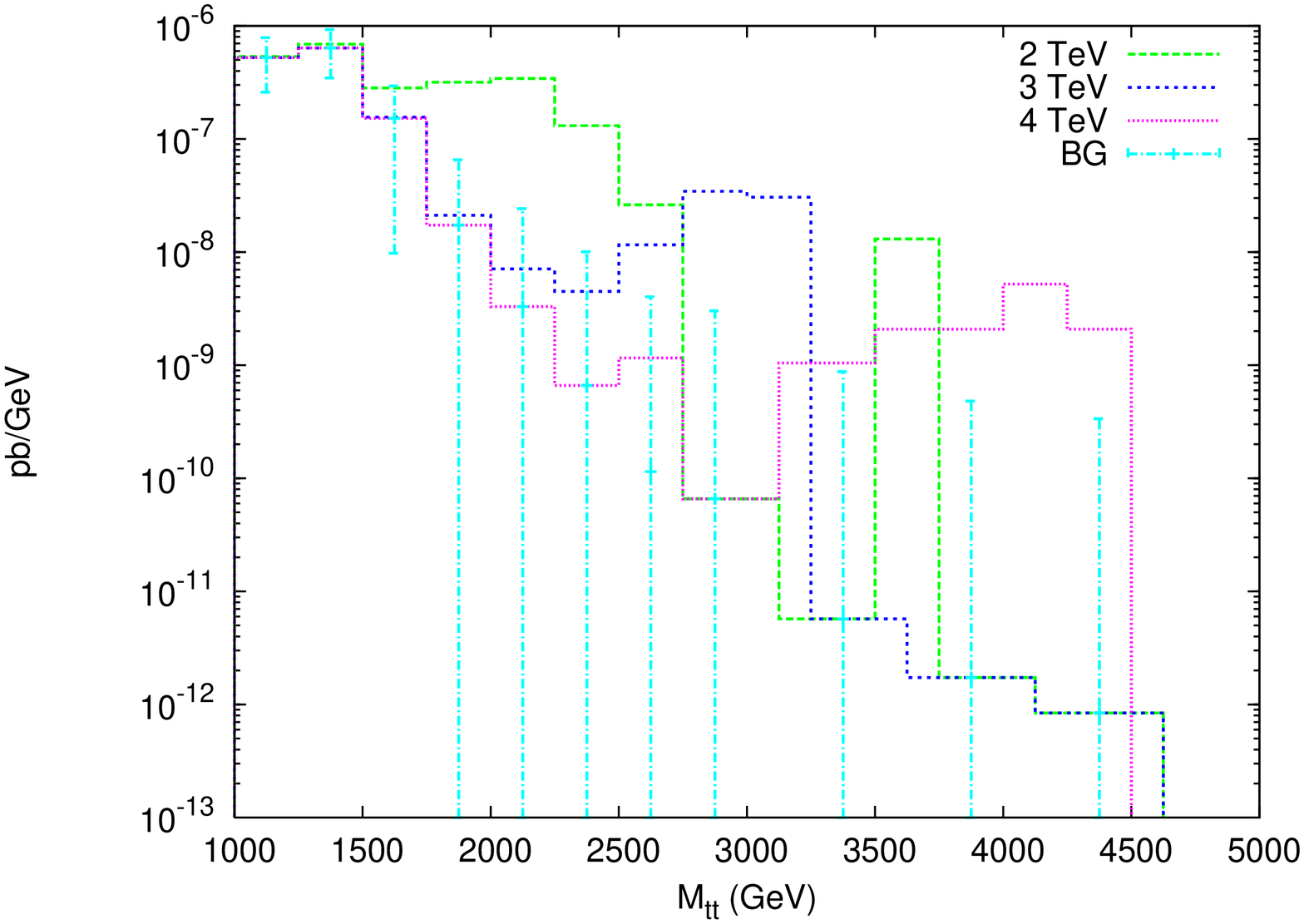}
\end{center}
\caption{\label{fig:liso}Invariant mass distribution for events
where at least 1 lepton is isolated. A $p_T$ cut of $500$ GeV on one
of the tops has been used.}
\end{figure}

To study  this collimation, we have generated SM
$t\bar t$ events in PYTHIA \cite{Sjostrand:2006za}. In
Figs.~\ref{fig:nobj} and \ref{fig:nobj-pt}, we discuss possible cuts
relevant to top quark identification as a function of energy and top
$p_T$. In these figures we have imposed a lower $p_T$ cut and a
window of invariant masses and then find the fraction of events
which have a certain number of isolated objects (jets, leptons, or
$\not\!\!p_T$). We use an isolation cut of $\Delta R =
\sqrt{\Delta\eta^2 + \Delta\phi^2} > 0.4$, the smallest cone size
generally considered for jets. We also show the fraction of events
where at least one top is completely collimated, {\it i.e.} all
three decay products are within $0.4$ of each other.

For resonances heavier than 3 TeV, there is always a large portion,
$> 90 \%$, of the events with at least one top jet highly collimated. If these cannot be distinguished from QCD jets,  the full QCD multi-jet cross-section would be the
background for fully hadronic events, and the $W+$jets cross-section
for the leptonic modes. This is in addition to other reducible
backgrounds. Carefully chosen cuts may reduce these. Nonetheless, it is clear that an
effective strategy for identifying highly collimated tops and
distinguishing them from QCD jets will be  essential to identifying $t \bar{t}$ events in this kinematical regime.

We also see that for $M_{KK}>3(4)$ TeV, a significant portion
$30\%$($70 \%$) of  events will have only two well isolated objects.
 Clearly here the problem of identifying highly
collimated top quarks is the most serious and the most critical.
In order to not be swamped by the QCD dijet background,   new ways of identifying these highly collimated
top jets will be essential.  For $M_{KK}<2$
TeV, we expect the conventional method of reconstructing two tops
from six distinct objects will work, at least for the majority of the
events.

Notice that top collimation, within a particular window of $m_{t
  \bar{t}}$, the collimation depends strongly on $p_T$. This
stems from the difference in boosting a central and a forward top.
For a forward top, it is generically easier for the decay product to
have a large $\phi$ separation. At the same time, the boost also has
less effect on $\Delta \eta$, since it is an invariant if the top is
moving along  the $z$-direction. Therefore, forward top quarks tend
to have more separated decay products. At the same time, within an
invariant mass window, a $p_T$ cut tends to force the tops to be
more central. This seems to indicate that we should focus on more
forward tops. However, notice that SM $t \bar{t}$ background is even
more forward, and a strong $p_T$ cut is usually necessary to
suppress those backgrounds. Therefore, we have to deal with the more
severe collimation problem for the central tops.

Lepton isolation is also important for a variety of reasons. It points
toward a class of cleaner events. It is also important in
measurements such as top helicity, as we discuss in Section 4.
 In
Fig. \ref{fig:leps} we plot the fraction of events with an isolated
lepton, with the same cuts as used above. We see that, if $M_{KK}>3$
TeV,  only a few $\%$ of the signal events  have an isolated lepton.
  Note also the difference between signal and background,
which is due to the fact that the signal events are more central
than the background events. Of course, even with an isolated lepton,
we still have to deal with the fact that the other side of the same
event will have the collimation problem alluded to above. Since the
fraction of events with an isolated lepton is different between the
signal and background, we can not estimate the relative rates simply
by scaling Fig. \ref{fig:mtt}. To understand the relative rates
between signal and background, in Fig. \ref{fig:liso} we show the
invariant mass distribution of events with a separated lepton. We
see that, with a $p_T$ cut of $500$ GeV on one of the tops the
signal does indeed dominate over the background, but that the event
rate is becoming low.

Over the mass range $2$ TeV $<M_{KK}<$ $3$ TeV, there are a wide
variety of event configurations featuring from $2$ to $5$ separated
objects with different portions depending on mass and all these cases need to be considered. Each of them will be a combination of partial
reconstruction and/or identification of collimated top jets. Each of
them, depending on which object is isolated,  will have to  include
different backgrounds. A complete study, certainly important and
interesting, is beyond the scope of this paper. When there are more than two jets, we  expect the reconstruction efficiency to be somewhat
better than the two objects case due to the additional handle of the extra
separated objects, but a quantitative assessment is required.

We now comment on  possible techniques for identifying the highly collimated
"top-jets". If jets can be identified as originating from top quarks,
even without explicitly separated decay products,
one can nonetheless identify the resonance structure in the
$t \bar{t}$ invariant mass distribution as well as the angular
distributions of the tops, which carries the spin information of the
initial KK-gluon state.

One route is to always demand some partial
separation between some of the objects, such as a $b$ and a lepton.
Indeed, this is probably what one has to do in order to measure the
top quark helicity, and we have seen that this will likely work very
well for KK masses $< 2$ TeV, and will probably work for masses
$\lesssim 3$ TeV. It is probably somewhat easier to do independent
lepton identification for muons than for electrons, depending on
ECAL performance, and the understanding of the EM energy content of
high $p_T$ jets. At the same time, the question still remains how to
reconstruct or identify the other side (hadronic) as a potential top
candidate. Moreover, even demanding only a single isolated lepton
without any cuts on the other decay products, we see that the event
rate takes a big hit when $M_{g^{(1)}} \gtrsim 2$ TeV.

Most efficient top tagging algorithms will  require the existence of
a b-tag. However, tagging an energetic b-quark with additional
objects close to it will be challenging so the tagging rate at high
$p_T$ is still uncertain. In particular one should worry that the
high boost of the $b$ will cause the opening angle of the $b$ decay
products to close beyond the resolution of the detector. It is also
possible that the merging of the $b$-jet with the other top decay
products will cause the $b$-tagging efficiency for the signal to be
lower than that for the background.

Other strategies for identifying top quarks will exploit the differences between a top jet and a normal jet. One distinguishing feature of the top jet is its large
invariant mass. Some massive jet algorithms, for example,
similar to those used in identifying hadronic Ws in $WW$ scattering
\cite{Wjmass}, might be useful for  identifying a top jet.
However,  QCD processes could produce massive jets
with off-shell partons and hard radiations. For example,   the
average jet mass, conventionally defined, for jets with high $p_T$
peaks at around $15\%$ of the jet $p_T$ \cite{Campbell:2006wx} so a full
background study is required.

One additional important aspect of top-jets might be sub-structure,
which can potentially be probed by using more detailed features of
events. For example, sub-clusters at the tracking level could be
particularly useful since the inner detectors usually have finer
granularity than calorimeters. Variables one might want to study
include sub-cluster with a common impact parameter, or more than two
subclusters. The presence of high $p_T$ leptons merged with the jets
may also prove useful. A jet where a large fraction of the energy is
contained in one muon, for example, is unusual for QCD jets.

With the extreme importance of the top signal in mind, it is
worthwhile to study these questions in much greater detail. Any such
strategy will have to be optimized, probably for every particular
mass region, to gain maximal efficiency and reduce fake rate.
Different strategies will involve different backgrounds which could
fake the signal. As such a study is  beyond the scope of this
paper, we will demonstrate the main features of KK-gluon
phenomenology by parameterizing the degrading of the signal with a
range of fake rates and leave detailed studies of the suggested
methods to the future.

\subsection{Extracting the KK-gluon signal}

In this section, we study how well the signal can be extracted over
background. We present the signal in Fig. \ref{fig:mjj}. Note an interesting feature of the high mass resonance signal. The events are almost pure $s$-channel resonant
production yet most of the events from the high-mass
resonances show up at low invariant mass, many widths away from the
resonance. This is due to the steeply falling PDFs in the region of
the resonance which compensates the suppression from high virtuality.

\begin{figure}
\begin{center}
\includegraphics[angle=270,scale=0.5]{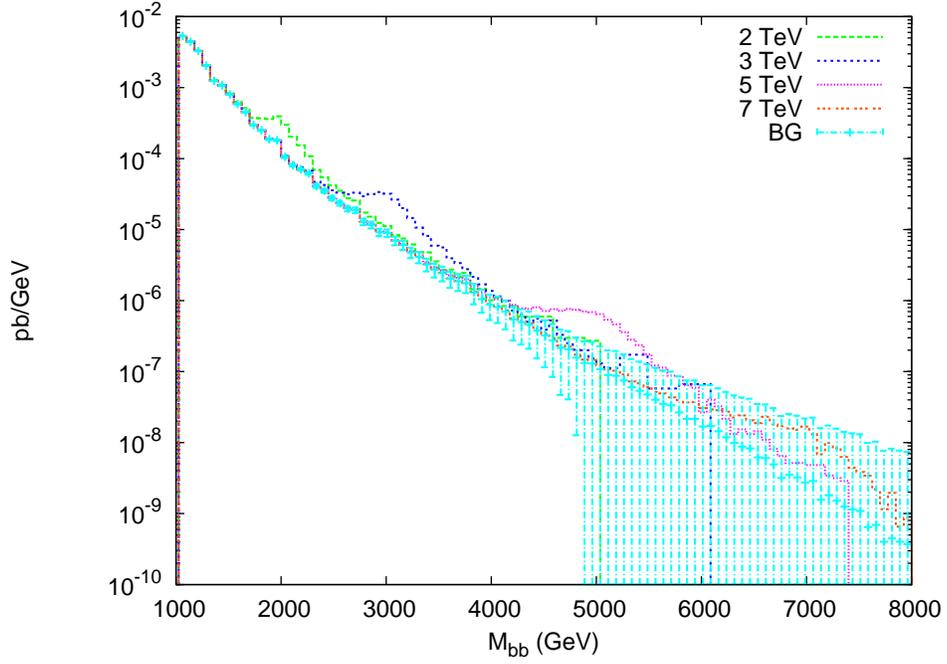}
\end{center}
\caption{\label{fig:mbb} Same as in the previous figure, but
requiring 1 $b$-tag. All signal curves have also been added to the
background curve. }
\end{figure}

We first consider the extent to which  $b$-tagging  could help. If we
assume $50\%$ efficiency and a $1\%$ fake rate \cite{atlasTDR}, both
for the signal and the background and demand one $b$-tag we get the
distributions shown in Fig. \ref{fig:mbb}. These numbers for
$b$-tagging performance are almost certain to be wrong at high
$p_T$, but this provides an optimistic estimate in the absence of a
detailed study in the range of $p_T$ considered here. We see that there is some possibility that the
resonance will be observable just by requiring a single $b$-tag if
it is highly efficient, as assumed here.

Notice that we can also conclude from this plot that the $b_L$ final
states are unlikely to be useful for discovering the resonance. The
rate is reduced by a factor of 17 relative to top quark production.
Since the background in Figure \ref{fig:mbb} that remains after
b-tagging is predominantly $b$ $\bar{b}$, we can see that the signal
will be swamped by the background.

In general, to identify the energetic top quarks,  new techniques
will be required. Developing a proper top-tagging algorithm is
beyond the scope of this paper. As stated in the previous
subsection, we will instead investigate the behavior of the signal
with different assumptions for the performance of such an algorithm.

\begin{figure}
\begin{center}
\includegraphics[angle=270,scale=0.3]{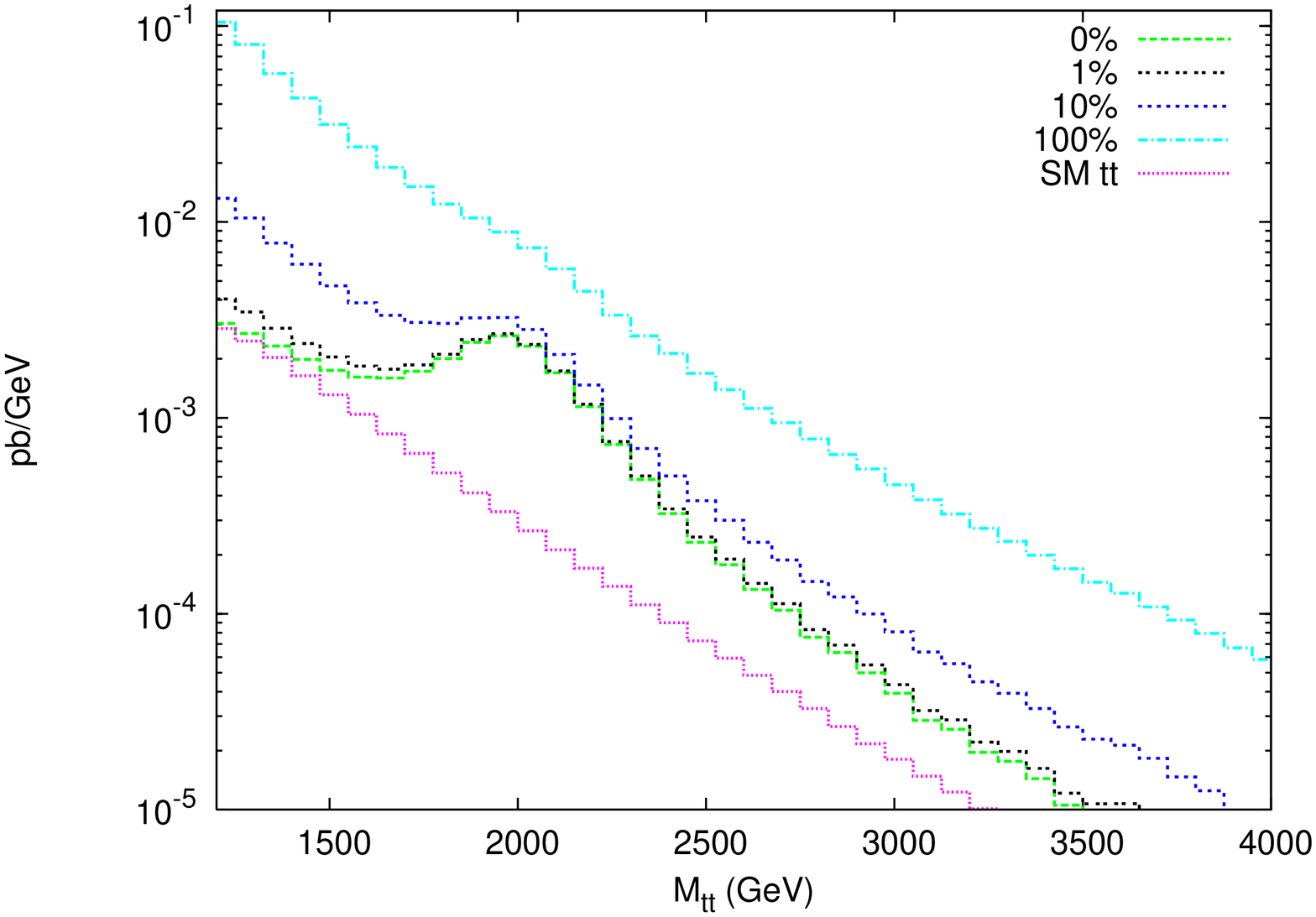}
\includegraphics[angle=270,scale=0.3]{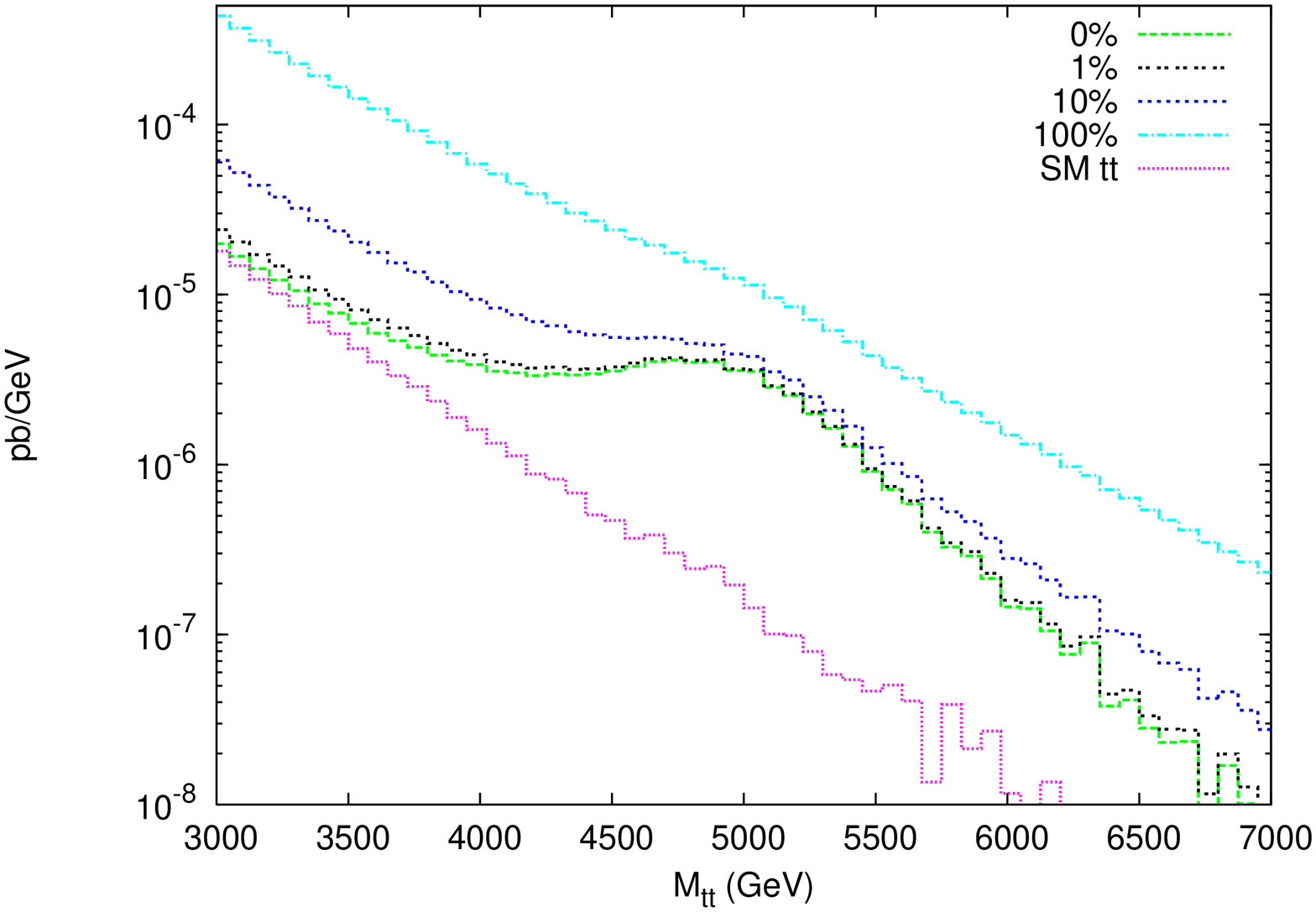}
\end{center}
\caption{\label{fig:dilute}$t\bar t$ invariant mass distribution for signal
plus some background
assuming various values of the $t$-tagging fake rate for a resonance
at 2 (left) and 5 (right) TeV.}
\end{figure}

Every top identification strategy will introduce fakes from its own
background. The strategy we consider with two object final states
will inevitably have to contend with, among other backgrounds, fakes
from the QCD dijet and $b \bar{b}$ backgrounds. Given the size of
the dijet and $b\bar b$ rates, the important feature of top
identification will be more background rejection than signal
efficiency. We have simulated this by assuming a rate for dijet
events faking $t\bar t$ events. The resulting invariant mass
distributions for resonances of 2 and 5 TeV are plotted in
Fig.~\ref{fig:dilute}.

From the plot, we conclude that extraction of the signal will
require a background rejection of about a factor of 10. Notice that
since we do not include a reduction factor on the $t\bar{t}$ (both
signal and background), the result presented here will be similar to
a more inclusive algorithm which is optimized to include as much $t
\bar{t}$ as possible. More refined balance may have to be sought
between purity and efficiency in a completely realistic study.

For a 5 TeV resonance, a rejection factor of 10 for the dijet events
produces a signal to background ratio of $S/B \sim 1$. Comparing
Figs. 3 and 8 we can see that this is true for all KK masses we have
considered. So a rejection factor of 10 will allow discovery as long
as the luminosity is high enough to produce sufficiently many
events. For $100$ fb$^{-1}$, a resonance at 6 TeV will produce about
10 events after signal cuts. While it may be possible to discover a
resonance with so few events, it would take a careful detector-level
study to determine if this is indeed possible. Hence, our
conservative estimate is that resonances with masses up to 5 TeV
could be discovered at the LHC. The luminosity upgraded SLHC will
likely extend this reach.

\subsection{Spin measurement}

After finding a resonance, it will be important to determine its properties. One important question is the spin of the resonance.   Since the KK-gluons are produced via the process $q
\bar{q} \rightarrow g^{(1)}$, we expect its decay products will
reproduce the familiar $1+\cos^2 \theta$ distribution in the rest
frame of the KK-gluon.

\begin{figure}
\begin{center}
\includegraphics[angle=270,scale=0.5]{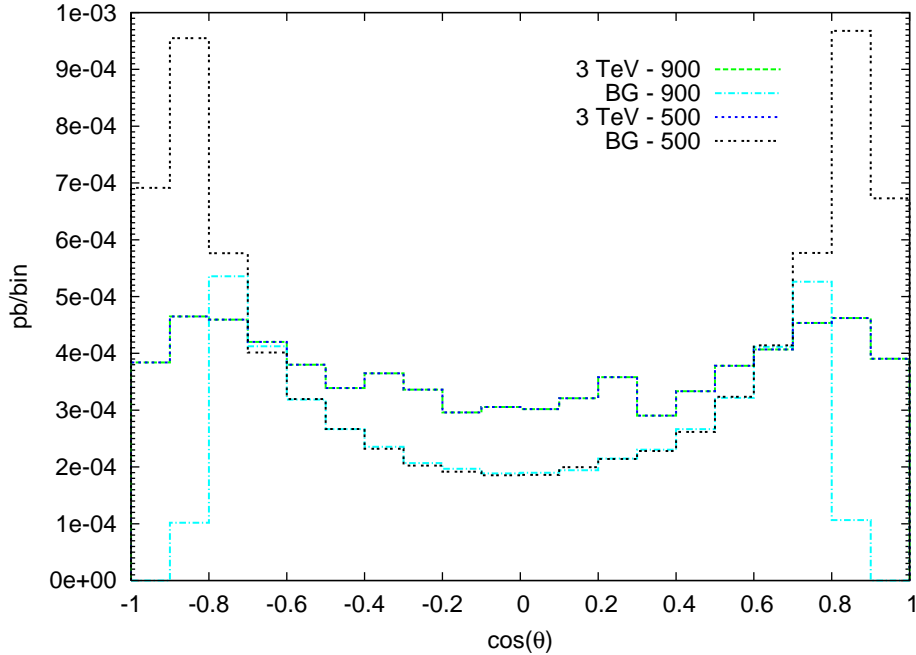}
\end{center}
\caption{\label{fig:costheta}Distribution of $\cos\theta$, the decay
angle in the hard-scattering CM frame , from a $3$
  TeV KK-gluon resonance. We show signal and background
for two different choices of $p_T$ cuts.}
\end{figure}

Finding the spin, i.e., probing the $(1+\cos^2  \theta)$ structure
of the distribution,  requires making use of the full angular
acceptance of the central detector. Here, we used a 3 TeV KK-gluon
as an example, though the  results are similar for other
masses. We require that the top decay products are all
restricted to the range $|\eta| < 2.5$, and that the invariant mass
is in a narrow window around the resonance, $2800 < m_{tt} < 3500$.
Since the background is steeply falling, we use a window that is
asymmetric around $m_{tt}$, with the lower bound being closer to
$m_{tt}$ than the upper bound. This should enhance the signal to
background ratio.
Additionally, because the masses are large we can further enhance
this ratio by imposing a strong $p_T$ cut on either the
reconstructed top or the top-jet. We plot the angular distributions
in Fig \ref{fig:costheta} for two different cuts on  $p_T$,
either $500$ or $900$ GeV. We see that a suitably strong $p_T$ cut
has very small effect on the signal. This is expected since the
$p_T$s of the tops are peaked around $0.5 \ M_{KK}$. On the other
hand, such a cut is quite effective in suppressing the background.
Crucial information about a $(1+\cos^2  \theta)$ distribution comes
from the more forward part of the distribution. However, the
background is also strongly peaked forward. We see that an
appropriate $p_T$ cut could effectively reduce this part of the
background.

\section{Partial compositeness of $t_R$}

As has been known for more than a decade,
spin-correlation  \cite{topspin-th}  in $t\bar{t}$ production provides an important
probe of the structure of the Standard Model. It has since become an
important subject for experimental study \cite{topspin-fnal}, and
will be explored in much more detail at the LHC \cite{topspin-atlas}.

With the $t \bar{t}$ final states produced from the decay of some
heavy new physics resonances well beyond twice the top mass, spin
polarization measurements of the tops produced in the decay should
be another powerful tool for probing new states, such  the KK-gluon,
which is likely to be  mostly coupled to right-handed tops. In this
section we will demonstrate the
 significance of the spin polarization measurement\footnote{Other measurements of the
  chirality of the coupling come from forward backward asymmetry measurements and
  left-right  asymmetries. These measurements give important Z-pole
  observables. However, the forward-backward asymmetry measurement
 requires both the product vertex
  and the decay vertex of the resonance to be chiral, which
is not generically satisfied \cite{RSmodel} for the KK-gluon since
both the left and right handed quarks of the first two generations
are localized towards the Planck brane. Clearly a left-right asymmetry
 measurement is impossible at a hadron collider.}.

Top quarks produced from a particular vertex will be of some
definite chirality depending on the  vertex.
Such a state will generically be a mixture of helicity eigenstates
since chiral symmetry is broken by the top quark mass. However, if
very energetic top quarks are produced, they
will be approximately in a helicity eigenstate, (to the
order of $\mathcal{O}(m_t/E)$). Therefore, such top quarks could be
thought of as a good chirality analyzers since their helicities carry
the information of initial chirality of the coupling.

Extraction of the helicity of the top quark follows from a well-known
procedure \cite{topspin-th,topspin-fnal,topspin-atlas}. We consider
here the leptonic decay mode $t \rightarrow b W^+ \rightarrow b
\ell^+ \nu $. In the rest frame of the top quark, the direction of
top three-momentum $\mathbf{\hat{k}}$ defines the axis of polarization,
i.e., the helicity of the top quark. We first consider the case in
which the top is right-handed.  Due to the $V-A$ nature of the
Standard Model weak interaction, bottom quarks from top decay must
be left handed. Depending on whether the $b_L$ goes along or
opposite to the direction of the top, angular momentum conservation
requires that the $W$ will be polarized either along $\mathbf{\hat{k}}$
($|J=1, m=1>$)\footnote{We have used $\mathbf{\hat{k}}$ to define
the polarization axis for $W$. } or longitudinally ($|J=1, m=0>$),
respectively. Angular momentum conservation and $V-A$ again require
$\ell^+$ will always be correlated state of W polarization, either
goes along the direction of its polarization for $|J=1, m=1>$ state,
or goes perpendicular to the polarization axis for $|J=1, m=0>$
state. Combining these arguments, we conclude that $\ell^+$ will
always prefer to go along (or perpendicular) to $\mathbf{\hat{k}}$,
but not opposite to it\footnote{A very nice discussion of this can
be found in \cite{Tait:1999ze}.}.

In the case of a left-handed polarized top quark, a similar argument
leads to the conclusion that $\ell^+$ will prefer to go along the
opposite (or perpendicular) to $\mathbf{\hat{k}}$. A careful
treatment shows that

\begin{gather}
\frac{d\Gamma}{\Gamma d\cos\phi_i} = \frac{1}{2}(1+\alpha_i
\cos\phi_i)
\end{gather}
where $i$ labels the decay product, and $\alpha_i$ is a constant.
For the lepton (or down-type jet) $\alpha_\ell = 1$ at leading
order, while for the $b$ and neutrino (or up-type jet) $\alpha_i
\approx -0.44$; NLO corrections to the $\alpha_i$ are known to be of
order 1\% \cite{topspin-atlas}.

Therefore, by measuring the angular distribution of the $\ell^+$ in
the rest frame of the $t$ with respect to $\mathbf{\hat{k}}$, we
will be able to establish the initial polarization of the top
produced. Therefore, in the process of $g^{(1)} \rightarrow t
\bar{t}$, due to the correlation between helicity and initial
chirality, measurement of such angular correlation will establish
the chirality of the $g^{(1)}-t-\bar{t}$ coupling.

\begin{figure}
\begin{center}
\includegraphics[angle=270,scale=0.5]{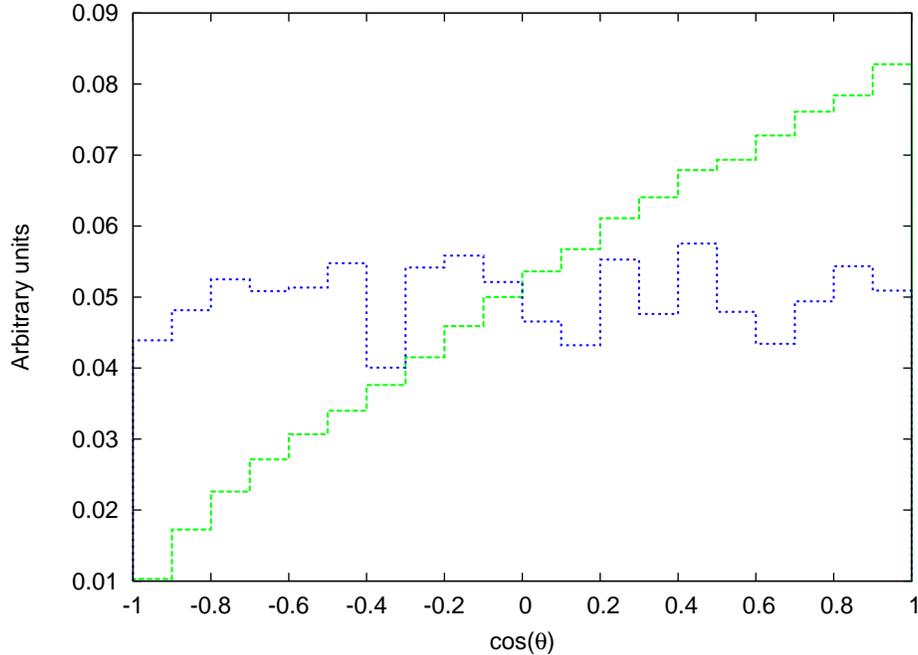}
\end{center}
\caption{\label{fig:coslt}Normalized distribution $\cos(\phi)$,
defined in the text, for a 3 TeV resonance (blue, dotted), and for
QCD top production (green, dashed).}
\end{figure}

We demonstrate such a measurement in the simple setup of
\cite{RSmodel}, where we expect this coupling is dominantly
right-handed. For this we use the full matrix element for $pp \to
t\bar t \to e^+\nu_e\,b\bar b\,jj$, generated from MADGRAPH. Figure
\ref{fig:coslt} shows the angular distribution for a 3 TeV resonance
(without background) and for QCD top production. With an isolated
lepton it should be simple to extract the top polarization. This
tells us that such a measurement will be possible for KK masses up
to $\sim 2$ to $3$ TeV where the lepton isolation is difficult.

\begin{figure}
\begin{center}
\includegraphics[angle=270,scale=0.5]{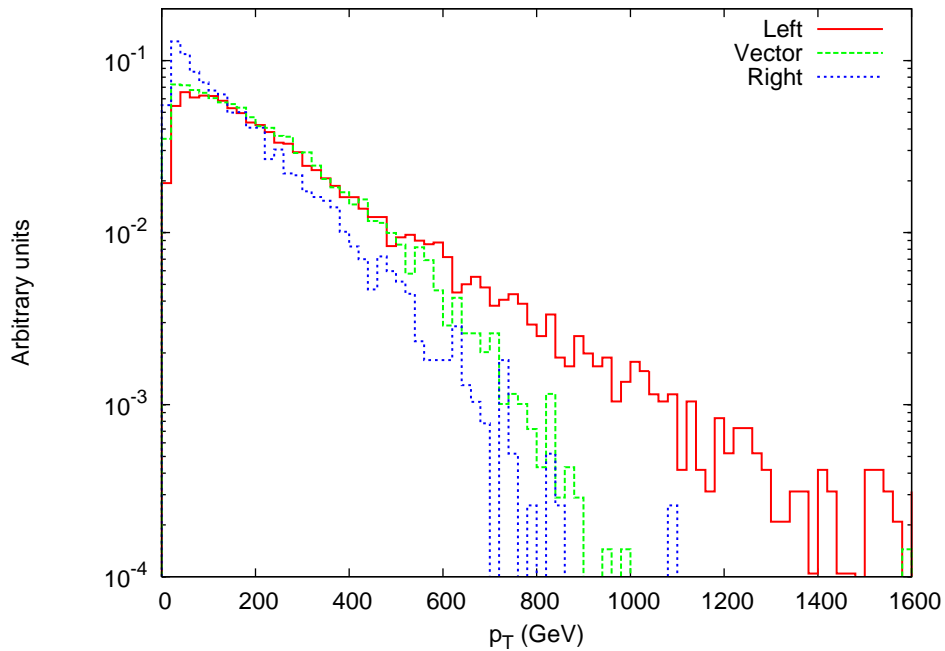}
\end{center}
\caption{\label{fig:plt}Distribution of lepton $p_T$ from the
process $pp\to t\bar t \to e^+ \nu_e b\bar b jj$ for three
resonances. One coupled like the KK gluon described in the text
``Right''), one where the couplings to left and right handed tops
have been reversed (``Left'') and one coupled equally to left and
right (``Vector''). }
\end{figure}

Another possibility is to use the $p_T$ spectrum of the lepton as a
proxy for the top polarization, as was done in studies of the $W$
helicity in top decays at the Tevatron \cite{w-helicity}. To
illustrate this we looked at a 2 TeV resonance with 3 different sets
of couplings. One is coupled in the same way as the rest of this
paper; one where the left and right handed couplings to tops are
switched; and finally one where the coupling is vector like. In all
cases the sizes of the couplings were chosen to keep the width
constant. We plotted the normalized lepton $p_T$ spectra in Fig
\ref{fig:plt}. This technique could, in principle, be used to study
higher mass resonances since it does not depend on the lepton being
isolated.

\section{Conclusions}

In this paper, we have studied the $t \bar{t}$ final state as a
discovery mode and probe of the KK-gluon properties in the RS
scenario where Standrad Model fields are propagating in the bulk. We pointed out
the for the energetic tops decaying from resonances with mass $\geq
2$ TeV, conventional top reconstruction methods will be limited,
despite the high production rate. Without some top jet identification,
the signal will be swamped simply by the QCD dijet background.

We showed that $b$-tagging information will help  reduce the
background. However
in order to efficiently identify the energetic tops,  more detailed
information from the top jets will be required. Implementation of
such a strategy will certainly require detailed knowledge of the top
``jet'' structure and detector response to it. Given the importance
of such a signal, it merits more detailed study. At
the same time, any such method could not completely eliminate the
fakes from background, such as QCD dijets and $b \bar{b}$ pairs. In
the current study, we studied the question  of how much discrimination
power against such fakes is required in order to have a
significant signal and take advantage of the big
cross section for KK-gluon production and subsequent decay to the
$t$ $\bar{t}$ final state.  In a study in which we parameterize the fake rate,
we found that a rejection power of a factor of 10 is required in order
to achieve this goal.

We also considered the angular correlations of the decay
products of the top quarks to deduce the top quark helicity.  Decaying
from some heavy resonance,
the top quarks are very boosted. Therefore, the helicity is highly
correlated with the chiral structure of the decay vertex. Such a
measurement is crucial to probe the profiles of the SM fermions in
the bulk. In the simple setup \cite{RSmodel}, the KK-gluon couples
dominantly to right-handed top quarks. We demonstrated ways to
establish the right-handed nature of such coupling, through
conventional correlation observable in the top rest frame, as well as
a $p_T^{\ell}$ observable.

Additionally, we remark that top quarks are most likely to be
strongly coupled to any model of the electroweak sector and
therefore identifying top quark jets could be one of the more
important experimental goals. In particular tht top jet final state
should be an important  signature for any strongly interacting new
physics for which top quarks are somewhat special, such as composite
models or topcolor models. Although we have focused on KK-gluons and
using \cite{RSmodel} as an example, we expect many lesson we have
started to learn here will have a wide range of applicability.

Finally, we comment on, Ref.~\cite{Agashe:2006hk} which appeared
recently and has overlap with our study. We note that there are
several differences in the approach, scope, and results. Most
noticeably, they have focused on a particular method of identify
hadronic tops without taking advantage of the top jet properties.
Hence, strong cuts are imposed, which could underestimate the
discovery potential. They used helicity of the $t_R$ as a way
sharpening the signal. While it is certainly a useful approach, we
remark that such a property is model dependent. In principle, it
should come as a result of the measurement.

\section{Acknowledgments}

We would like to thank T. Han, Joey Huston, Greg Landsburg, Tom
LeCompte, Liam Fitzpatrick, Jared Kaplan, Zack Sullivan, Tim Tait,
and Chris Tully for useful discussions. B.L. is supported by the
Department of Energy contract DE-AC02-06CH11357. The work of L.W. is
supported by the National Science Foundation under Grant No. 0243680
and the Department of Energy under grant DE-FG02-90ER40542.  Any
opinions, findings, and conclusions or recommendations expressed in
this material are those of the author(s) and do not necessarily
reflect the views of the National Science Foundation.

\appendix

\section{Formalisms}
\label{formalisms}

We are particularly interested in a model where all Standard Model
(SM) fields except the Higgs propagate in the entire 5-d spacetime,
and will be primarily concerned with the gluon and colored fermion
fields. The pure gauge action for the gluons is the
standard\footnote{The indices $M,N$ run over $(0,1,2,3,5)$, while
$\mu,\nu$ run over $(01,2,3)$; $a$ is a gauge index.}
\begin{gather}
S_{\rm gluon} = \int d^5x \sqrt{g}
-\frac{1}{4}F_{\mu\nu}^{a}F^{\mu\nu\, a}.
\end{gather}
We employ the gauge $A_5 = 0$, do a standard KK reduction
\begin{gather}
A^a_\mu (x,y) = \sum_n A^{a(n)}_\mu
\frac{\chi^{(n)}(y)}{\sqrt{r_c}},
\end{gather}
and solve for the wavefunctions $\chi^{(n)}$. The solution for the
wavefunction of the $n$-th KK mode is\cite{RSmatterinbulk}
\begin{gather}
\chi_A^{(n)}= \frac{e^{k\pi\phi}}{N_A^{(n)}}\left[
J_1(m_A^{(n)}e^{k\pi\phi}/k)+
\alpha_A^{(n)}Y_1(m_A^{(n)}e^{k\pi\phi}/k)\right].
\end{gather}
Here $J_1$ and $Y_1$ are the standard Bessel functions of the first
and second kind, $m_A^{(n)}$ is the mass of the $n$th KK mode, and
$N_A^{(n)}$ is a normalization factor. Note that for an unbroken
gauge group there is a zero mode which is flat in the extra
dimension, $\chi^{(0)} = 1/\sqrt{2\pi}$.

For the fermion sector, note that in five dimensions there is no
chirality. We can thus always write down a mass term for each 5d
fermion field. Hence the fermion action for a single species,
$\Psi$, is
\begin{gather}
S_{\rm fermion} \int d^5x \bar\Psi i \not\!\! D \Psi +
m_{\Psi}\bar\Psi\Psi.
\end{gather}
Since $\Psi$ is Dirac, we can write it as $\Psi = (\zeta\ \xi)^\top$
where $\zeta$ and $\xi$ are Majorana spinors. We can write the mass
as $m_{\Psi} = \nu_{\Psi}k \epsilon(y)$, where $\epsilon(y)$, which
is $1$ for $y>0$ and $-1$ for $y<0$, is responsible for making the
mass term even under the orbifolding symmetry. The orbifolding
projects out the zero mode of one chirality. The wavefunction of the
remaining zero mode is\cite{RSflavor}
\begin{gather}
\xi_{\Psi}^{(0)} = \frac{e^{\nu_{\Psi} k \pi \phi}}{N_f^{(0)}},
\end{gather}
where again $N_f^{(0)}$ is a normalization factor. This wavefunction
is exponentially peaked to toward the UV brane for $\nu < -1/2$ and
toward the IR for $\nu > -1/2$. These zero modes correspond to the
observed SM fermions, and will acquire masses from the Higgs vev in
the normal way. Since the Higgs is localized to the IR brane the
effective Yukawa coupling for a SM fermion species $f$ is
\begin{gather}
\lambda_{f} = \lambda_0 \xi_{L}(\pi)\xi_{R}(\pi),
\end{gather}
where $\lambda_0$ is the dimensionless 5d Yukawa coupling. For
fermions with one or both components with $\nu < -1/2$ this will be
exponentially suppressed. Choosing different ${\mathcal O}(1)$
values for the $\nu_{\Psi}$ allows the entire fermion mass hierarchy
to be generated\cite{RSflavor}.

\end{document}